\documentclass[11pt,a4paper]{article}

\usepackage[margin=1in]{geometry}
\usepackage{amsmath,amssymb}
\usepackage{graphicx}
\usepackage{booktabs}
\usepackage{bm}
\usepackage[numbers,super,sort&compress]{natbib}
\usepackage{hyperref}
\usepackage{lineno}
\usepackage{setspace}
\usepackage{xcolor}

\newcommand{\hh}{\mathbf{h}}
\newcommand{\uu}{\mathbf{u}}
\newcommand{\ssm}{\mathbf{s}}
\newcommand{\xx}{\mathbf{x}}
\newcommand{\yy}{\mathbf{y}}
\newcommand{\zz}{\mathbf{z}}
\newcommand{\Am}{\mathbf{A}}
\newcommand{\Sm}{\mathbf{S}}
\newcommand{\PP}{\mathbf{P}}
\newcommand{\RR}{\mathbf{R}}
\newcommand{\XX}{\mathbf{X}}
\newcommand{\Mm}{\mathbf{M}}
\newcommand{\qq}{\mathbf{q}}
\newcommand{\pp}{\mathbf{p}}

\newcommand{\cc}{\mathbf{c}}
\newcommand{\pphi}{\bm{\phi}}

\hypersetup{colorlinks=true,linkcolor=blue,citecolor=blue,urlcolor=blue}

\begin{document}

\title{Hypervectors from Optical Disorder: Programmable Encoding and Optical Inference for Hyperdimensional Computing}

\author{Takuya Iwata$^{1}$, Namthip Srisuthep$^{2}$, and Satoshi Sunada$^{2,*}$}

\date{}
\maketitle

\begin{center}
$^{1}$Graduate School of Natural Science and Technology, Kanazawa University,\\
Kakuma-machi, Kanazawa, Ishikawa 920-1192, Japan \\[2pt]
$^{2}$Faculty of Mechanical Engineering, Institute of Science and Engineering,\\
Kanazawa University, Kakuma-machi, Kanazawa, Ishikawa 920-1192, Japan \\[2pt]
$^{*}$Corresponding author: \texttt{sunada@se.kanazawa-u.ac.jp}
\end{center}

\vspace{1em}

\begin{abstract}
\noindent
Hyperdimensional computing (HDC) is a computing framework that represents information as high-dimensional pseudorandom vectors called hypervectors (HVs), enabling learning and inference through simple algebraic operations. The HV dimensionality provides computational capacity and error robustness, but the required high-dimensional randomness must be stored or regenerated, imposing a memory--computation tradeoff. Here, we address this tradeoff by physically embodying the randomness required for HV generation in the static disorder of a scattering medium. 
Specifically, a silicon photonic circuit combined with the scattering medium generates high-dimensional HVs from a small number of input values in a single optical shot. Detector-side encoding programs HV correlations for continuous and categorical representations. The resulting HVs reproduce key statistical and compositional properties of ideal i.i.d.\ random HVs, including pairwise similarity statistics, associative-memory capacity, and factorization capacity. We further demonstrate an optically assisted inference path using the generated HVs and optical similarity evaluation.
\end{abstract}

\section*{Introduction}
The growth of artificial-intelligence workloads has intensified the demand for computing architectures that deliver high throughput at low energy cost. Photonic computing is a compelling candidate, exploiting the intrinsic parallelism and low energy dissipation of light to perform linear-algebraic operations more efficiently than electronic counterparts \cite{feldmann2021,shen2017,McMahon:2023,Miller:2017,Hamerly:2019,Chen_Science2025,Meyer:2026,Sved:2026,Ou:2025,Luan:2026,Zhao:2025,Bente:2025}. However, photonic neural networks largely inherit the electronic paradigm of layer-wise nonlinear activation and gradient-based weight updates, requiring frequent electro-optical conversions that can erode photonic advantages in latency and energy efficiency \cite{shastri2021,Nahmias:2020}. Moreover, analog noise, device nonidealities, and accumulated errors further limit system precision and scalability. These limitations motivate computing models and architectures that minimize data conversions and nonlinear operations while intrinsically tolerating analog imperfections.

Hyperdimensional computing (HDC), also known as vector symbolic architectures \cite{kanerva2009,Plate:2003,Kanerva:1996,ge2020}, is a promising framework for addressing these challenges. HDC uses distributed representations based on ultrahigh-dimensional pseudorandom vectors, called hypervectors (HVs), and processes them through simple operations, including bundling (component-wise addition), binding (component-wise multiplication), permutation, and similarity comparison [Fig.~\ref{fig:concept}(a)]. Consequently, HDC is robust to noise and supports low-precision computation \cite{rahimi2016,thomas2021,Kleyko:2022a,HernandezCano:2021}, while the simple and low-complexity vector operations of HDC facilitate hardware implementations, including electronic in-memory computing \cite{karunaratne2020,Wu:2018,Langenegger:2023} and photonic computing \cite{Fayza:2026,Najafi:2025,LiuMa:2025}. Beyond these hardware-friendly features, HDC-based architectures have demonstrated capabilities such as visual analogical reasoning \cite{Hersche:2023}, memory-augmented few-shot classification \cite{Karunaratne:2021}, and factorization of holographic representations for disentanglement \cite{Langenegger:2023}. HDC has also been applied to diverse practical tasks \cite{Rahimi:2016b,Imani:HDNA:2018,Zhuowen_BioHD:2022,Mitrokhin:2019,Yue:2026,Kitagawa:24,Kleyko:2023,Schlegel2022,Arockiaraj:2026,Kleyko:2022b}.

The computational capability of HDC is strongly tied to the HV dimensionality $D$. As $D$ increases, independently generated random HVs become increasingly quasi-orthogonal, with their residual similarity decreasing as $\sim 1/\sqrt{D}$. This high-dimensional representation supports distributed encoding and can improve associative capacity, inter-class separability, and robustness to noise \cite{kanerva2009,thomas2021,Kleyko:2022b}. Dimensionality is therefore a fundamental computational resource of HDC, but also a hardware challenge, because the high-dimensional randomness required to construct HVs must be supplied from somewhere. 

Conventional implementations store this randomness either in an item memory holding one HV per symbol or in a random projection matrix for continuous features [Fig.~\ref{fig:concept}(b)]. In both cases, the memory requirement grows with $D$ and with the number of represented items or input features \cite{Aygun2023_learning}. A common strategy to alleviate this burden is to regenerate HVs on the fly, trading memory for additional computation \cite{Schmuck:2019,Kleyko:2022c}. Considerable effort has been devoted to easing this memory--computation tradeoff through algorithmic and hardware-efficient designs \cite{Thomas_streaming2023,Tang:2026,Saeidi:2026}, but these approaches still face challenges such as increased hardware complexity and deviations from ideal statistical independence \cite{Saeidi:2026}. This raises the question of whether the randomness required for HVs can be supplied without either storing it or recomputing it digitally, while preserving the statistical properties needed for HDC.

Here, we address this question with an optical approach in which the randomness required for HV generation is supplied by the static disorder of a scattering medium (SM). Using a silicon photonic circuit coupled with the SM, we generate a $D$-component HV in a single optical shot from $M$ input values, with $M \ll D$, thereby reducing the digital memory and computation required to supply the high-dimensional randomness.

Although optical random projections through SMs have been used as feature maps in other computing paradigms \cite{saade2016,rafayelyan2020,Sunada:20,Wang_Light2025}, here we exploit them for programmable HV encoding: the correlations among the generated HVs can be tuned to the input data and computational task through simple post-processing (detector-side encoding), without modifying the optical system. We show that the measured HVs exhibit the key statistical and compositional properties of ideal random HVs required for HDC. Furthermore, we experimentally demonstrate an optically assisted HDC inference path in which element-wise multiplication for similarity evaluation is implemented optically, advancing optical HDC beyond numerical simulations \cite{Fayza:2026,Najafi:2025,LiuMa:2025} and beyond digital HDC processing of optical signals \cite{Kitagawa:24,Yue:2026}.

\section*{Results}
\subsection*{Physical hypervector generation}

\begin{figure}[t]
  \centering
  \includegraphics[width=\textwidth]{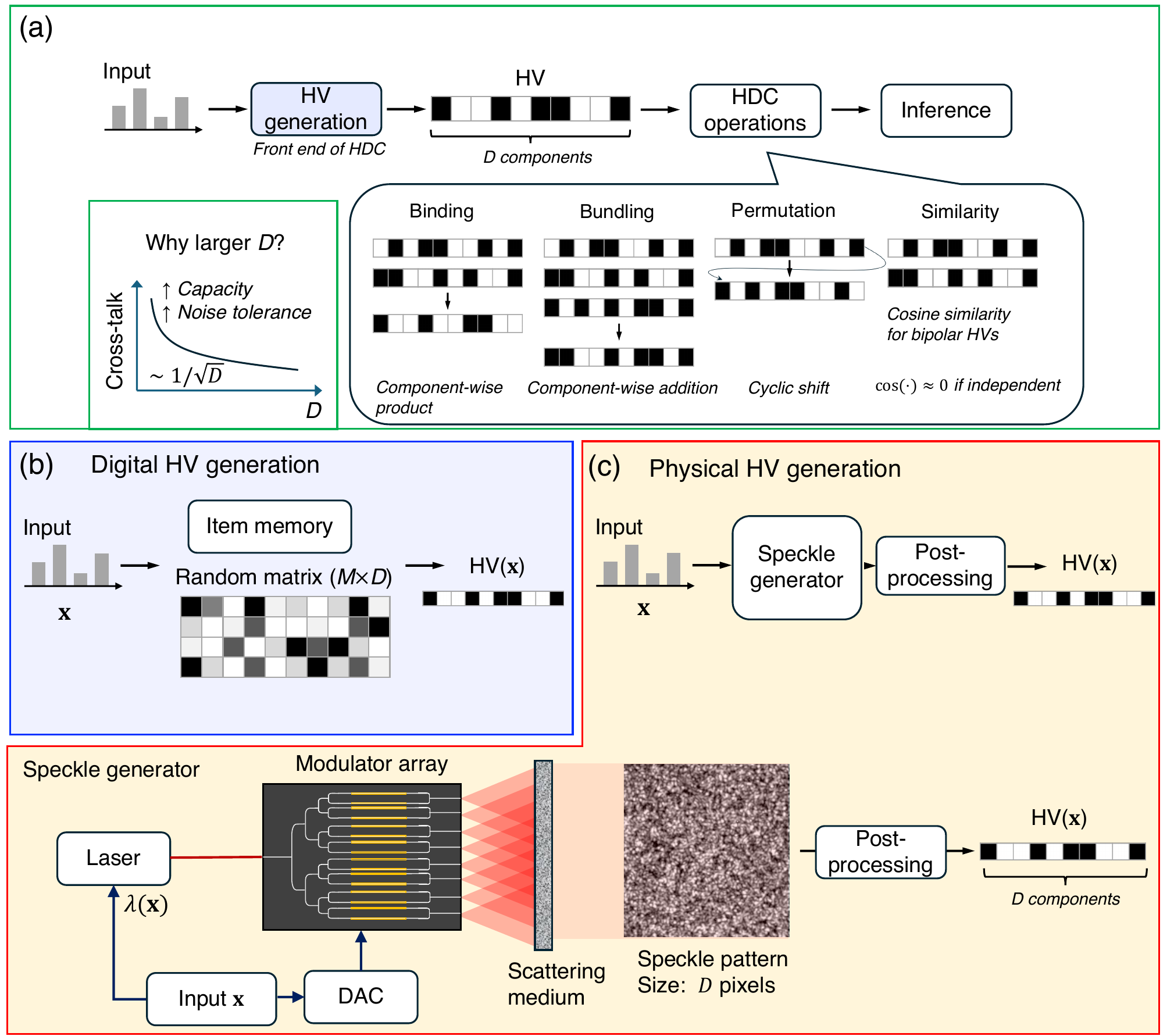}
\caption{\textbf{Physical hypervector generation.} \textbf{(a)} HDC represents data as $D$-dimensional hypervectors (HVs) and operates on them with four closed-form primitives. Increasing $D$ improves quasi-orthogonality and computational capacity. \textbf{(b)} Conventional HV generation requires hardware resources that scale with $D$. \textbf{(c)} The proposed generator uses a photonic integrated circuit with $M'$ modulators and a passive scattering medium to generate an HV comprising $D$ sampled components from an input $\xx$, with optional wavelength encoding $\lambda(\xx)$.}
\label{fig:concept}
\end{figure}

Figure~\ref{fig:concept}(c) illustrates the proposed optoelectronic HV generator, which consists of a speckle generator followed by a post-processing stage. Coherent laser light is modulated by an array of $M'$ Mach--Zehnder modulators (MZMs) on a silicon photonic integrated circuit (PIC), driven through a multi-channel digital-to-analog converter (DAC) (Supplementary Note 1). Each MZM carries two independently driven phase shifters and therefore encodes at most two components of the input feature vector, $\xx=[x_1,\dots,x_M]\in\mathbb{R}^{M}$, so that $M'=\lceil M/2\rceil$; the $m'$-th modulator converts $x_{2m'-1}$ and $x_{2m'}$ into the complex optical amplitude, $z_{m'} \;=\; \bigl[\exp(\mathrm{i}\phi(x_{2m'-1}))+\exp(\mathrm{i}\phi(x_{2m'}))\bigr]/\sqrt{2}$, where $\phi(x) \in \mathbb{R}$ denotes the phase shift driven by the input $x$. The $M'$ modulated fields are relayed to a passive SM and co-illuminate it (Methods), where multipath interference distributes the input across the detection plane. The speckle pattern at the detection plane is represented as
\begin{equation}
  \Sm \;=\; \left|\Am\zz \right|^2 \in\mathbb{R}_{\geq 0}^{D},
  \qquad
  \Am=\PP_2\RR\PP_1 \in\mathbb{C}^{D\times M'},
  \label{eq:speckle}
\end{equation}
where $\zz=[z_1,\dots,z_{M'}]^{\!\top}\in\mathbb{C}^{M'}$ collects the MZM outputs, $\PP_1$ and $\PP_2$ are the propagation matrices before and after the SM, and $\RR$ is the transfer matrix of the SM. Optionally, the laser wavelength $\lambda$ can also be tuned according to the input $\xx$ [Fig.~\ref{fig:concept}(c)], such that the transfer matrix becomes wavelength dependent, $\Am\to\Am(\lambda)$. This provides an additional addressing channel. The post-processing stage in Fig.~\ref{fig:concept}(c) completes the generator by converting the captured intensity pattern into binary or bipolar HVs. 

Equation~\eqref{eq:speckle} shows that the SM physically performs a random projection of the $M'$-dimensional modulated field onto $D$ sampled components. 
For a strongly scattering medium, the elements of $\Am$ approach circular complex Gaussian statistics \cite{Goodman:2007}, although neighboring detector coordinates remain correlated within a speckle grain (Supplementary Note~2). In our system, $M=14$ input components are encoded by $M'=7$ modulators, while a $640\times512$ sensor region provides up to $D_{\max}=327{,}680$ sampled components, exceeding $M$ by more than four orders of magnitude.

Throughout this work, $D$ denotes the number of sampled detector coordinates used as HV components. It differs from the approximate number $N_{\rm sp}$ of independently resolved speckle grains, which we estimate as $N_{\rm sp}\simeq2\times10^{4}$ for the present system (Supplementary Note 2). Denser sampling can therefore increase $D$ without increasing the underlying optical degrees of freedom. The requirement for HDC is not that the HVs carry $D$ independent bits, but that they behave as i.i.d.\ random vectors under binding, bundling and similarity. We accordingly assess the HDC-relevant statistical dimensionality from the measured similarity and capacity of the generated HVs, which are shown below to closely follow the scaling of ideal i.i.d.\ HVs.

\subsection*{Programmable correlation structure of the generated HVs}

HDC uses different types of HVs for different kinds of data. Categorical symbols require mutually quasi-orthogonal HVs to prevent interference, whereas continuous quantities require level HVs whose similarity varies smoothly with the difference in value. Conventional implementations typically construct these HVs separately, for example using an item memory for categorical symbols and a dedicated set of level vectors for continuous values. Here, both types are generated by the same physical process and selected through post-processing at the detection stage.

\begin{figure}[t]
  \centering
  \includegraphics[width=\textwidth]{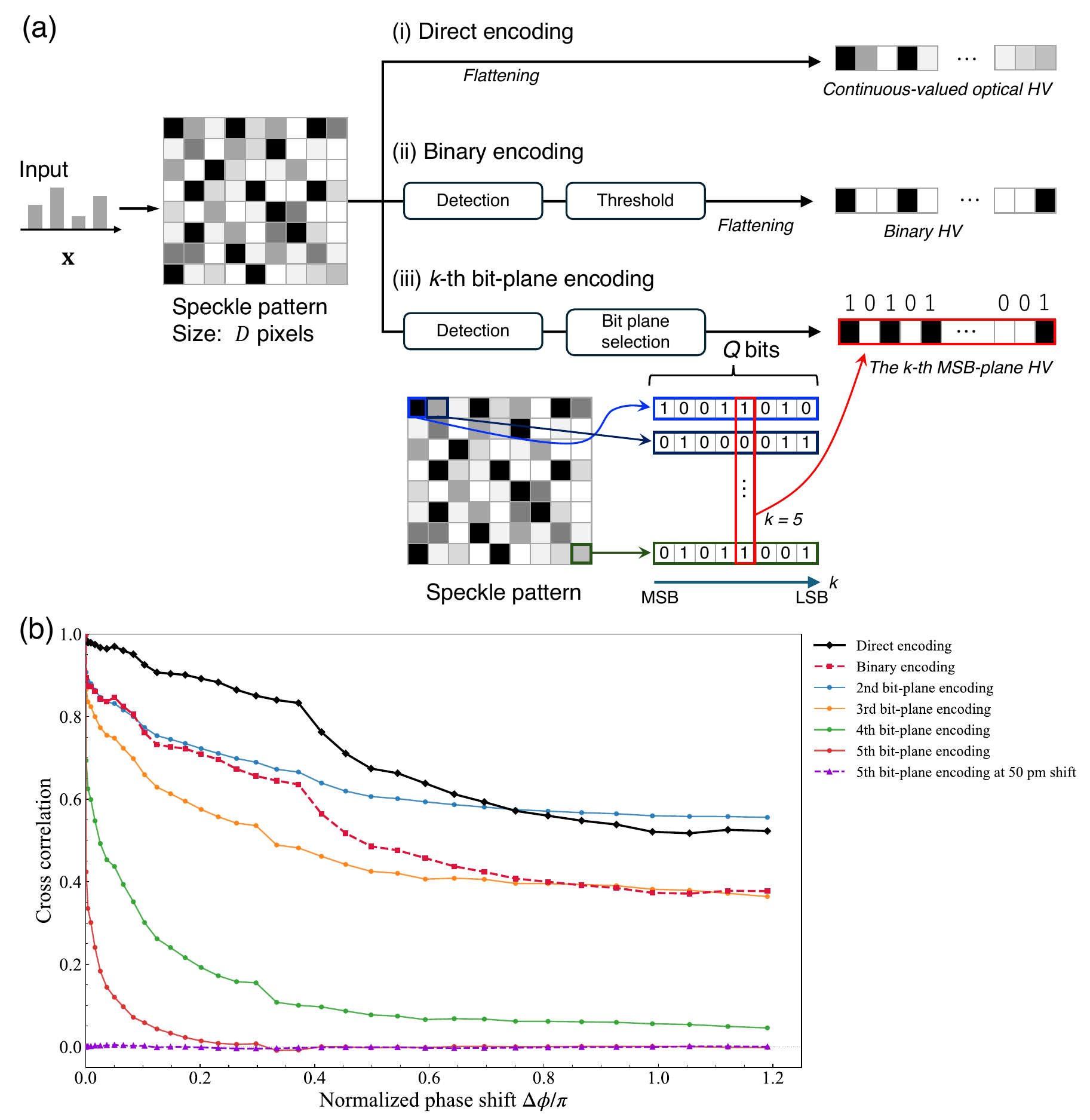}
\caption{\textbf{Programmable HV encoding.} \textbf{(a)} Three encoding methods using a single speckle pattern: direct, binary, and $k$-th bit-plane encoding. 
\textbf{(b)} Pearson cross correlation between a reference HV and phase-shifted HVs versus input phase shift $\Delta\phi = |\pphi|$ relative to the reference. Binary and bit-plane HVs are mapped to bipolar form.}
\label{fig:encoding}
\end{figure}

Figure~\ref{fig:encoding}(a) shows the encodings supported by a single speckle pattern. In \emph{direct encoding}, the speckle pattern $\Sm$ is used as it is, without thresholding or quantization: it gives a continuous-valued optical HV that can be fed directly into the optical inference described below. In \emph{binary encoding}, $\Sm$ is thresholded at its spatial mean, giving a one-bit-per-component HV that is compact to store and inexpensive to bundle and compare. These HVs can be used for the encoding of continuous input values because they preserve input similarity. In \emph{$k$-th bit-plane encoding}, the detected intensity is quantized to $Q$ bits per pixel and only the $k$-th most significant bit plane is retained, giving a binary HV whose correlation structure is controlled by the single integer $k$. Thus, the bit-plane encoding can be used for representing categorical symbols.

The mechanism of bit-plane encoding is straightforward. The $k$-th bit plane resolves differences of order $\langle \Sm\rangle 2^{-k}$, and therefore sets the scale in input space beyond which two inputs are treated as distinct. Upper planes follow the slowly varying intensity envelope and retain input similarity, whereas lower planes follow increasingly fine fluctuations, which approach independent random binary sequences~\cite{Uchida_inbook:2012}. 

Figure~\ref{fig:encoding}(b) shows the cross-correlation between HVs generated at MZM input phases $\pphi(\xx) = [\phi(x_1),\cdots,\phi(x_M)] \in \mathbb{R}^{M}$ and zero reference phases.
Direct and binary encodings retain a residual correlation floor of $\approx$0.5 and 0.4 due to the inhomogeneous illumination, and therefore cannot represent quasi-orthogonal symbols. 
In bit-plane encoding, increasing the bit-plane index $k$ progressively removes this floor, which reaches the quasi-orthogonality level at $k=5$ over most of the modulation range.
In our experiment, bit planes with $k \geq 7$ for $Q=12$ are increasingly affected by readout noise (Supplementary Note~2 and Fig.~S3). Thus, within the range $k \leq 6$, $k$ tunes the encoder from a similarity-preserving to an effectively orthogonal representation without changing the optical hardware.

The wavelength provides an additional address. For the $k=5$ bit-plane encoding, a $50\,\mathrm{pm}$ wavelength shift decorrelates the generated HVs at every input level [Fig.~\ref{fig:encoding}(b)]. The maximum number of generated quasi-orthogonal HVs is set by the product of the number of resolvable inputs and the number of resolvable wavelengths, and it grows exponentially with $M'=\lceil M/2\rceil$. Our rough estimate gives $O(10^{10})$ HVs for the present system (Supplementary Note 3). 

\subsection*{Dimensionality scaling of HDC capabilities}

The utility of HVs for compositional HDC depends on how their capabilities scale with dimensionality. We therefore examine three aspects in turn: concentration of measure, associative-memory capacity, and factorization capacity. In all cases, we use the HVs generated by $k$-th bit-plane encoding with $k=5$, which are mapped to bipolar HVs for the statistical characterizations. The measurements are compared with predictions for ideal i.i.d.\ random bipolar HVs.

\begin{figure}[t]
  \centering
  \includegraphics[width=\textwidth]{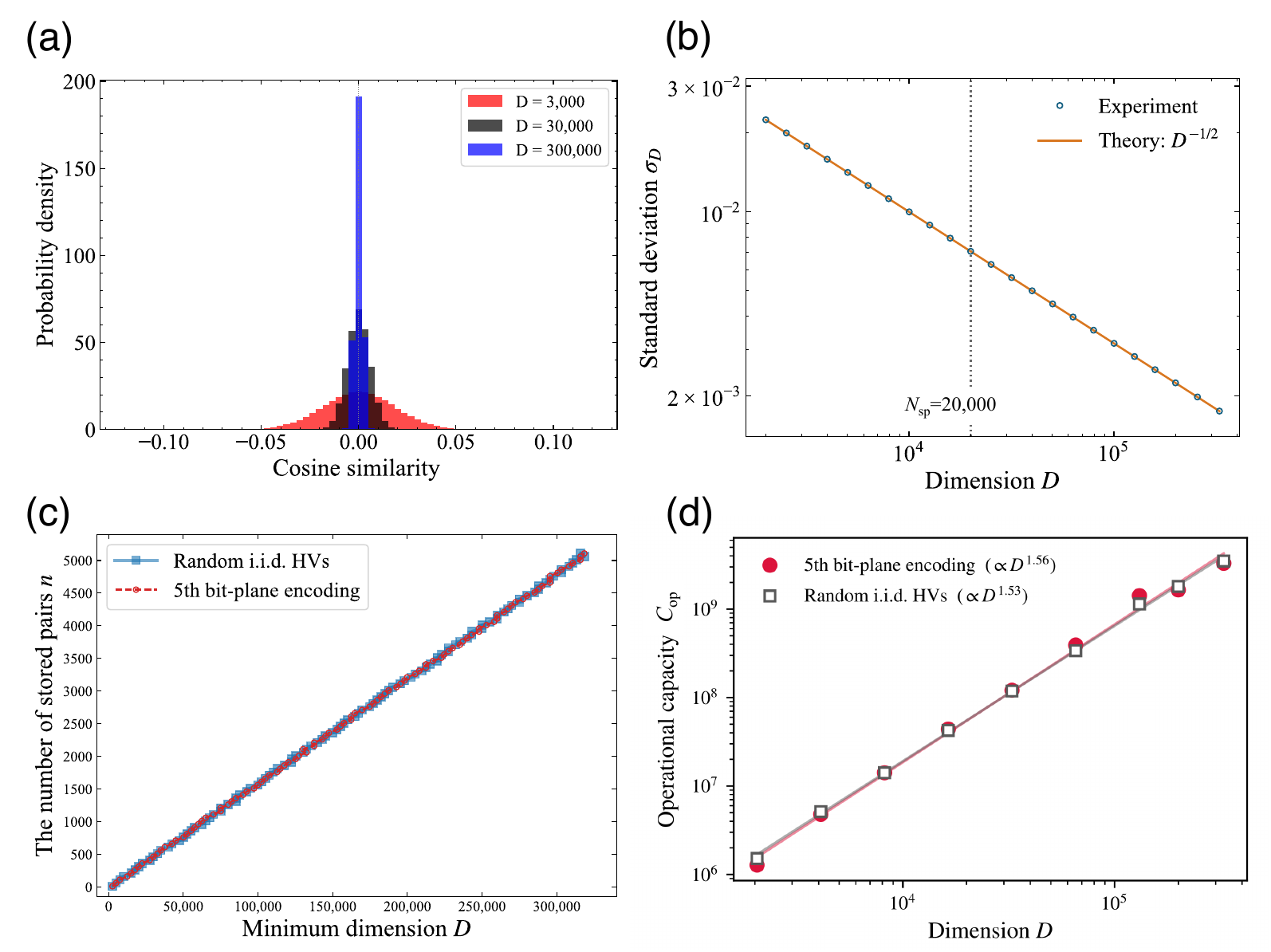}
\caption{\textbf{Generated HVs obey the scaling properties of HDC.} \textbf{(a)} Pairwise cosine-similarity distributions of bipolar HVs encoded at the $k=5$ bit plane. \textbf{(b)} Standard deviation $\sigma_D$ versus $D$ (markers), which closely follows the ideal prediction $D^{-1/2}$ (solid curve). For reference, the grain count $N_{\rm sp} \simeq 2\times 10^4$ is indicated by the dotted line. \textbf{(c)} Associative-memory capacity: the number of reliably stored key--value pairs $n$ versus minimum dimensionality $D$, compared with ideal i.i.d.\ HVs. \textbf{(d)} Factorization capacity for $F=3$. Codebooks use generated HVs (filled circles) or ideal i.i.d.\ bipolar HVs (open squares).}
\label{fig:scaling}
\end{figure}

\emph{Concentration of measure} [Fig.~\ref{fig:scaling}(a,b)].
Independently generated HVs become increasingly orthogonal as $D$ increases. Cosine similarity concentrates around zero for ideal i.i.d.\ random bipolar HVs. The measured distribution of pairwise cosine similarities narrows and concentrates around zero as $D$ grows over two orders of magnitude [Fig.~\ref{fig:scaling}(a)], and its standard deviation $\sigma_D$ follows the concentration-of-measure prediction $\sigma_D=D^{-1/2}$ $D\approx2\times10^{5}$ [Fig.~\ref{fig:scaling}(b)]. 

\emph{Associative-memory capacity} [Fig.~\ref{fig:scaling}(c)].
A defining feature of HDC is that many items can be superposed into a single HV of fixed length and still be read out individually. A standard test of this property is an associative memory: each key--value pair is bound by element-wise multiplication, all pairs are bundled into one memory HV $\Mm \in \mathbb{R}^{D}$, and a stored value is recovered by unbinding $\Mm$ with its key HV and comparing the result against the entire item memory \cite{Schlegel2022}. For each trial, we sample $2n$ distinct HVs from an item memory $\mathcal{V}=\{\hh_1,\ldots,\hh_N\}$ of generated HVs with $N \approx 10{,}000$ and assign them to $n$ key--value pairs $(\qq_i,\pp_i)$. We construct the bundled memory and retrieve each value as
\begin{equation}
\Mm=
\operatorname{sgn}\left(
\sum_{i=1}^{n}\qq_i\odot\pp_i
\right),
\qquad
\hat{\pp}_j=
\underset{\hh \in\mathcal V}{\mathrm{argmax}}\,
C(\Mm\odot\qq_j,\hh),
\label{eq:kv_bundle}
\end{equation}
where $C$ denotes cosine similarity and $\odot$ denotes element-wise multiplication. 
Because $\qq_j\odot\qq_j=\mathbf{1}$ for bipolar HVs, $\Mm\odot\qq_j$ is a noisy version of $\pp_j$, and cleanup against the item memory [Eq.~\eqref{eq:kv_bundle}] ideally returns $\hat{\pp}_j=\pp_j$; however, interference from the other stored pairs limits retrieval accuracy.
The associative-memory capacity increases with $D$ \cite{Plate:2003,Kleyko:2022b}. We quantify it as the smallest tested dimensionality $D$ at which the retrieval accuracy for $n$ stored pairs, averaged over 10 independently sampled sets of key--value pairs, reaches at least $99\%$. For fixed $N$, Fig.~\ref{fig:scaling}(c) shows that the number of reliably stored pairs $n$ grows approximately linearly with the required $D$, following the ideal scaling for random i.i.d.\ HVs over the full range examined.

\emph{Factorization capacity} [Fig.~\ref{fig:scaling}(d)].
Quasi-orthogonality and associative-memory capacity test pairwise similarity and retrieval from superpositions, respectively. A more stringent test is compositional search: given a product $\ssm=\hh^{(1)}_{a_1}\odot\hh^{(2)}_{a_2}\odot\cdots\odot\hh^{(F)}_{a_F}$ of HVs drawn from $F$ codebooks of $K$ entries each, where $\hh^{(f)}_{a_f}$ denotes the $a_f$-th entry of the $f$-th codebook and the indices $a_1,\dots,a_F$ are drawn uniformly at random, the task is to recover those indices. A resonator network performs this factorization in superposition, without enumerating the $K^{F}$ possible combinations \cite{Frady:2020,Kent:2020}. The largest search space that it can reliably resolve is quantified by the operational capacity $C_{\rm op}$, which grows super-linearly with the dimensionality $D$. This task therefore tests whether the generated HVs retain useful information through repeated binding, unbinding and cleanup. Codebooks were assembled entirely from generated HVs (Methods). Figure~\ref{fig:scaling}(d) shows the measured capacity $C_{\rm op}$ for $F=3$ as a function of $D$. The capacity follows $C_{\rm op}=10.9\,D^{1.56}$, while ideal i.i.d.\ bipolar HVs measured under identical conditions give $14.5\,D^{1.53}$; the ratio of the two fitted capacities remains between 0.94 and 1.10 over the full range. 

All three capabilities closely follow the ideal i.i.d.\ scaling up to the largest dimensionality examined, an order of magnitude above the grain count $N_{\rm sp}\simeq2\times10^{4}$ (Supplementary Note~2). The HDC-relevant dimensionality is therefore not capped by the number of independently resolved speckle grains.

\subsection*{Optical similarity-based inference with direct encoding}
HDC solves a classification task without iterative optimization. Of the three encodings in Fig.~\ref{fig:encoding}(a), bit-plane encoding supplies the quasi-orthogonal symbols characterized above; the experiments below instead use binary encoding for the class prototypes and direct encoding for the query. In our experiment, each training sample is converted into an HV, and the HVs of the same class are bundled into a single binary class HV,
\begin{equation}
  \cc_\ell = \Theta\!\left(\sum_{i\in\mathcal{T}_\ell}\hh^{\rm sp}_i
\right)
  \in\{0,1\}^{D},
  \label{eq:classify}
\end{equation}
where $\mathcal{T}_\ell$ is the training set of class $\ell$, $\Theta(\cdot)$ is the elementwise Heaviside step function, and $\hh^{\mathrm{sp}}_i=\mathrm{sgn}\!\left(\Sm(\xx_i)-\langle\Sm(\xx_i)\rangle\right)$ for the $i$-th training sample. $\langle\cdot\rangle$ is the spatial mean. 
For inference at an input $\xx$ of unknown class, the query HV is the speckle pattern $\Sm(\xx)$ itself under direct encoding; it is compared with every class HV, and the label of the most similar one is assigned,
\begin{equation}
  \hat{\ell} = \mathrm{argmax}_{\ell}\,\mathcal{S}_\ell,
  \qquad
  \mathcal{S}_\ell \equiv \cc_{\ell}^{\top}\Sm(\xx).
  \label{eq:infer}
\end{equation}
We take the inner product, rather than the cosine similarity, as the decision variable because it is the quantity measured optically. The two differ only by a normalization: the norm of the query is common to all classes, and the norm of a
class HV is fixed by the number of its active components, which differed by less
than $1\%$ between classes in our experiments. The same decision variable is used in the numerical reference (Methods).

\begin{figure}[t]
  \centering
  \includegraphics[width=\textwidth]{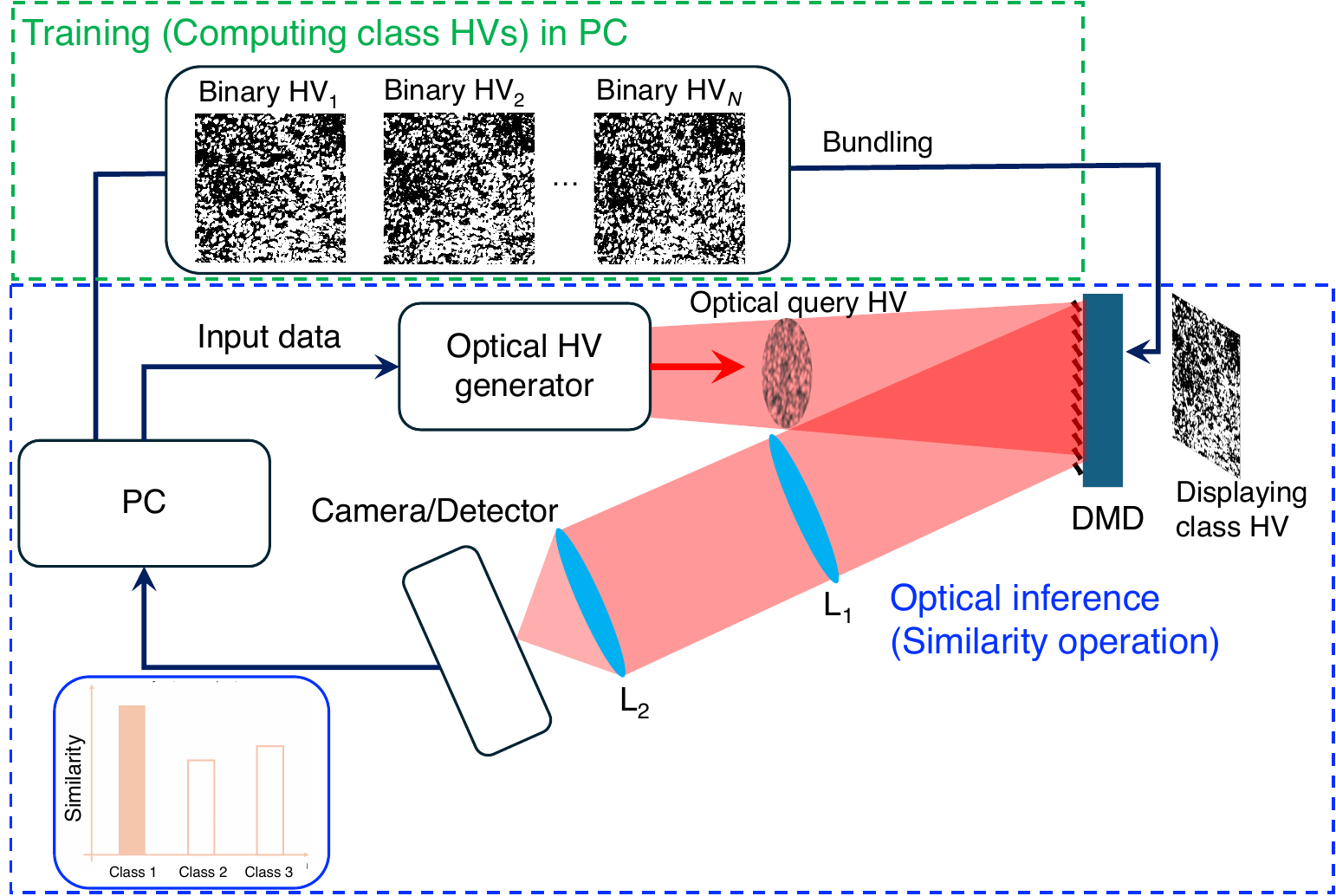}
\caption{\textbf{Optical implementation of the HDC inference data path.} Class HVs are computed once by bundling the binary HVs of the training samples (green box). During inference (blue box), the input drives the HV generator, and the resulting speckle pattern serves as the query HV. The query illuminates a DMD displaying one class HV, and lenses $L_1$ and $L_2$ collect the reflected light onto a detector/camera (also see Methods).}
\label{fig:inference}
\end{figure}

The class HVs (Eq.~\eqref{eq:classify}) are computed once on a computer by bundling the binary HVs of the training samples.
Figure~\ref{fig:inference} shows how this inference phase is implemented optically: the input drives the HV generator, and the resulting speckle field, corresponding to the query HV, illuminates a digital micromirror device (DMD). A DMD uses an array of tiltable micromirrors to rapidly display binary patterns, making it well suited for displaying the binary class HVs $\cc_\ell$ ($\ell=1,\dots,L$). In this experiment, one class HV is displayed at a time. The intensity distribution reflected by the DMD corresponds to the element-wise product of $\Sm(\xx)$ and $\cc_\ell$, and the total reflected power is proportional to $\mathcal{S}_\ell$. In the present setup, the sum over the $D$ components defining $\mathcal{S}_\ell$ is formed digitally over the registered pixels of the camera frame; an integrating photodetector placed in the collection path would perform it optically (Methods; Supplementary Note~4). Cycling the DMD through the $L$ class HVs yields $L$ such values, and the largest gives the predicted class.
 
\begin{figure}[t]
  \centering
  \includegraphics[width=\textwidth]{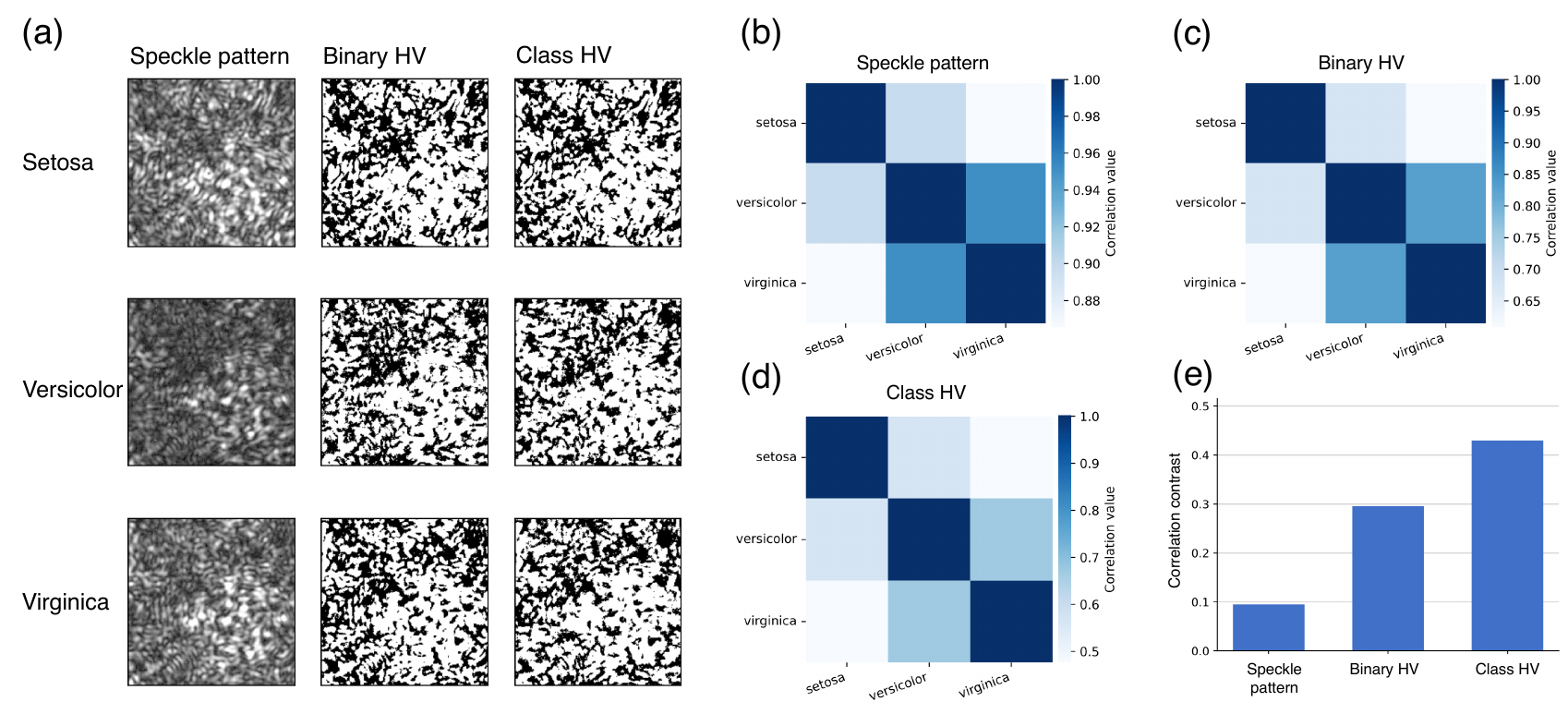}
\caption{\textbf{Optical HDC classification of the Iris dataset.} \textbf{(a)} Speckle patterns, binary HVs, and class HVs for one sample from each class. \textbf{(b--d)} Sample-averaged Pearson correlation matrices of the raw speckle patterns, binary HVs, and class HVs, respectively. \textbf{(e)} Correlation contrast $C_{\rm cont}$ at the three stages.}
\label{fig:iris}
\end{figure}

We first validate the pipeline on the Iris dataset \cite{Fisher_Iris_UCI} (3 classes, 4 features), chosen as a proof of principle rather than as a benchmark. Figure~\ref{fig:iris}(a) follows one sample of each class through the pipeline. The raw speckle patterns of different classes are strongly correlated, with off-diagonal correlations above $0.87$ [Fig.~\ref{fig:iris}(b)], because all three classes drive the same few input channels over a narrow range. Thresholding into binary HVs removes the common intensity offset and lowers the off-diagonal correlations [Fig.~\ref{fig:iris}(c)], and bundling into class HVs suppresses the sample-specific component that members of a class do not share, separating the classes further [Fig.~\ref{fig:iris}(d)]. Quantitatively, we measured the correlation contrast $C_{\rm cont}=1-\overline{\rho}_{\rm off}$, where $\overline{\rho}_{\rm off}$ denotes the mean correlation between samples of different classes. $C_{\rm cont}$ rises from $0.10$ for the raw speckle to $0.30$ after binary encoding and to $0.43$ after bundling [Fig.~\ref{fig:iris}(e)], showing that each HDC operation suppresses a further part of the component shared by all three classes. In our experiment, the optical system classifies the Iris test set with $85\%$ accuracy.

\begin{figure}[t]
  \centering
  \includegraphics[width=\textwidth]{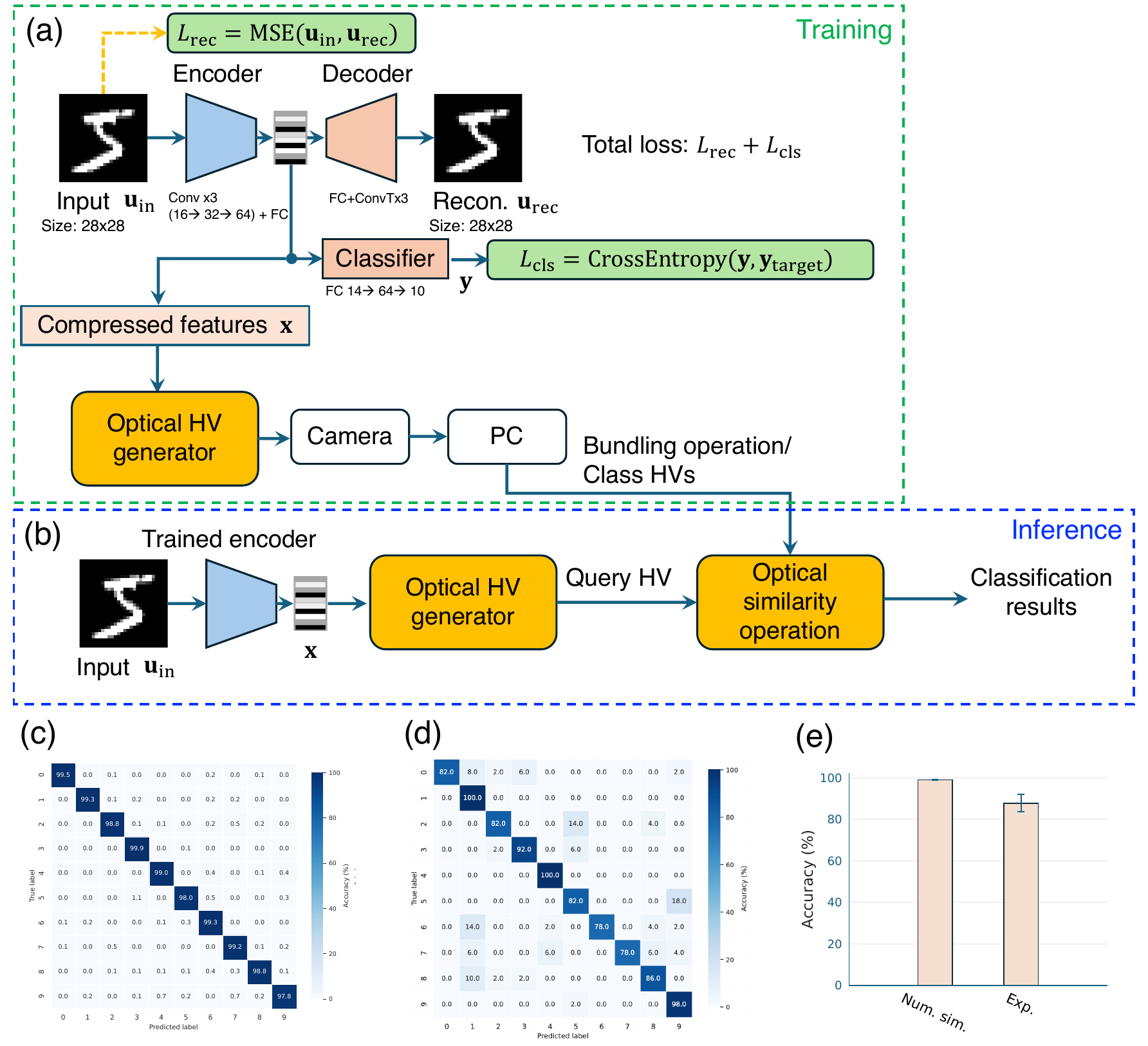}
\caption{\textbf{Optical HDC classification of MNIST with neural-network features.} \textbf{(a)} Training. A convolutional autoencoder with an auxiliary classifier maps each image $\uu_{\rm in}$ to a $14$-dimensional feature vector $\xx$, which drives the optical HV generator; binary HVs are bundled into class HVs. $\yy$ and $\yy_{\rm target}$ denote the output and target vectors, respectively. \textbf{(b)} Inference using the trained encoder and HV generator. \textbf{(c,d)} Confusion matrices for the numerical reference and optical system. \textbf{(e)} Classification accuracy.}
\label{fig:mnist}
\end{figure}

To further demonstrate the proposed optical inference scheme, we apply it to the MNIST dataset \cite{LeCun:1998}. MNIST is used here as a controlled benchmark for evaluating the physical implementation, rather than as a state-of-the-art image-classification benchmark. A convolutional autoencoder, trained jointly with reconstruction and classification losses, compresses each $28\times28$ image into an $M=14$-dimensional feature vector $\xx\in\mathbb{R}^M$ that matches the $M$ input channels of the HV generator [Fig.~\ref{fig:mnist}(a)]. At inference, only the encoder is retained: the feature vector drives the HV generator, and the resulting query HV is compared with the digitally prepared class HVs using the optical similarity operation [Fig.~\ref{fig:mnist}(b)]. The class HVs are constructed from only $100$ training images (Methods).

The numerical simulation achieves a high accuracy of $99\%$ [Fig.~\ref{fig:mnist}(c)], while the experimental system achieves $88 \pm 4.1\%$ [Fig.~\ref{fig:mnist}(d,e)]. The performance gap reflects additional errors introduced by the physical implementation, including analog fluctuations in detection and temporal drift, as well as differences between the idealized i.i.d.\ model of $\Am$ and the measured speckle. Thus, rather than serving as a benchmark of classification accuracy alone, this experiment demonstrates that the proposed HDC inference remains functional in the presence of hardware imperfections.

\section*{Discussion}

We have demonstrated an optical HV generator and combined it with an optical element-wise similarity operation to realize an optical inference data path. The randomness required to instantiate the HVs is embodied in the static disorder of an SM, rather than explicitly stored or digitally recomputed. This concept is analogous to electronic HV rematerialization \cite{Schmuck:2019}, but the optical approach represents $D$ components through a single passive medium without requiring a fabricated element for each component. The SM thus provides a physical substrate for HDC-compatible representations, rather than merely a high-dimensional feature mapper.

Neither the throughput nor the energy of the proof-of-concept system is fundamentally limited by the optics. In the present setup, the cooled image sensor dominates the total power consumption (Supplementary Note~4), while its frame rate and sequential DMD display limit the inference rate. These limitations arise from the readout architecture, not the computational principle. Because the class HVs are fixed after training, they could instead be displayed simultaneously on disjoint regions of a static DMD and read by $L$ photodetectors. This would yield all $L$ similarities in a single optical shot. 
The estimates of the energy and energy--delay product shown in Supplementary Note~4 indicate the potential advantages of extending the present architecture along these lines.

Three limitations remain. The first is that the optical advantage is confined to inference: learning and the bit-plane selection that yields quasi-orthogonal HVs are digital, so the compositional operations of HDC gain little from the optical substrate. Random phase-encoded holographic memory \cite{W-C_Su:2012} could provide a natural route to closing that gap, because the correspondence with HDC is direct \cite{Plate:2003}: writing object--reference interference, multiple exposure, and readout correspond naturally to binding, bundling, and similarity, respectively. Combining our optical HV generator with holographic memory would be an interesting future direction to address this issue.

Second, stability is a key concern due to our optical interferometric setup. Thermal drift or vibration can modify speckle patterns themselves, requiring training and inference to be completed within one correlation time. The weakest point in our system is the fiber relay: the modulated fields reach the SM through the seven independent cores of a multicore fiber (Methods), and the speckle pattern is their mutual interference, so a differential change in these seven path lengths changes it. With the temperature held within $\pm 0.5\,^{\circ}\mathrm{C}$, the pattern remains usable for about an hour, which limits the duration of an uninterrupted experiment. The relay is, however, a convenience of the proof-of-concept rather than a requirement: emitting the modulator outputs directly from the chip through grating couplers, either collimated onto the medium by a single lens \cite{Butow:2024} or launched into a multimode fiber that itself acts as the medium \cite{Hu:2023}, could place every channel on a rigid substrate and improve stability. 

Third, the input dimension is limited to $M=14$, restricting the richness of the encoded representations and potentially contributing to the reported benchmark accuracy. This limitation arises from the present chip, not the proposed scheme: $M$ determines the input-stage width, while $D$ and the optical similarity operation remain unchanged. Increasing $M'=\lceil M/2\rceil$ could exponentially expand the estimated number of addressable quasi-orthogonal HVs (Supplementary Note~3). Accuracy would also benefit from the retraining step standard in HDC, which is directly compatible here because it revises only the class prototypes displayed on the DMD.

All three limitations motivate integration. Combining a wider modulator array \cite{shen2017}, a disordered nanophotonic medium \cite{Zhao:2025}, and a multi-port detector array could place HV generation and similarity evaluation on a single substrate. The extent to which the estimated energy advantage survives such integration remains an open question.

\section*{Methods}

\subsection*{Optical setup}
The PIC comprises $M'$ thermo-optic MZMs. Each MZM carries two phase shifters, which are driven by an $M=2M'$-channel DAC. The voltage $V$ proportional to an input value $x$ is applied to the phase shifter; thus, the phase shift $\phi$ is approximately given by $\phi\propto x^2$. Coherent light around $\lambda=1550\,\mathrm{nm}$ from a wavelength-tunable laser (Santec, TSL-570) is coupled to the PIC input channel via a polarization-maintaining fiber and is split among the MZMs. The seven MZM output waveguides are coupled to the seven cores of a multicore fiber (MCF) (Fibercore Ltd., SSM-7C1500), which relays the modulated fields from the PIC to the free-space section of the setup. The seven beams emerging from the MCF output facet are collimated by a lens and co-illuminate a diffuser (Newport, 10DKIT-C3, diffusion angle: 60$^\circ$), which serves as the SM. The schematic of Fig.~\ref{fig:concept}(c) omits this fiber relay and collimating lens for clarity. The speckle pattern transmitted by the diffuser is imaged onto an InGaAs image sensor (BITRAN, CS-701IGA, $640\times512$ pixels, $5\,\mu\mathrm{m}\times5\,\mu\mathrm{m}$ pixel pitch, 12-bit, 133 fps). Photographs of the PIC and of the free-space section are shown in Supplementary Fig.~S1.

\subsection*{Generated HV library}
To characterize the ensemble of HVs that the generator produces, we recorded the HVs generated at different wavelengths under different input conditions. Figure~\ref{fig:encoding}(b) shows that, for $k=5$, HVs become mutually decorrelated for phase shifts $\Delta\phi\gtrsim0.36\pi$. At each wavelength, we therefore generated HVs for 17 input conditions, in which the phase shifts of all $M=14$ channels were set randomly such that any two conditions differed by $|\pphi^{(a)}-\pphi^{(b)}|>0.36\pi$. This procedure was repeated at 601 wavelengths from $1530$ to $1560\,\mathrm{nm}$ in $50\,\mathrm{pm}$ steps, yielding a library of $N=601\times17=10{,}217$ HVs. HVs of dimensionality $D$ are obtained by cropping each frame to $D$ sampled coordinates; when $D<D_{\max}$, non-overlapping crops of the same frame can serve as additional HVs. Throughout, the $k=5$ bit-plane HVs are converted to bipolar form, $\hh^{\rm bip}=2\hh-1\in\{-1,1\}^{D}$, before their statistics and capacities are evaluated.

\subsection*{Statistics of quasi-orthogonality and capacity}
The similarity distributions in Fig.~\ref{fig:scaling}(a,b) are computed from the $N$ HVs of the library, yielding $N(N-1)/2$ pairwise similarities for each $D$. For the associative-memory capacity in Fig.~\ref{fig:scaling}(c), $n$ key--value pairs are drawn from the same $N$ HVs and bound and bundled according to Eq.~\eqref{eq:kv_bundle}; we measured the minimum dimension $D$ required to retrieve $99\%$ of the $n$ pairs for a given $n$ and $N$. A full statistical characterization of the generated HVs, such as bit balance, pixel-wise means, and the behavior of the statistics under repeated binding, is given in Supplementary Note 5.

\subsection*{Factorization capacity}
$F$ codebooks $\XX^{(f)}=[\hh^{(f)}_1,\dots,\hh^{(f)}_K]\in\{-1,+1\}^{D\times K}$, $f=1,\dots,F$, were assembled from generated HVs, and composites $\ssm$ were formed with the indices $a_f$ drawn uniformly at random. The factors were recovered by a resonator network \cite{Frady:2020,Kent:2020} with the update
\begin{equation}
\hat{\hh}^{(f)}\;\leftarrow \; \mathrm{sgn}\!\left[
\XX^{(f)}\bigl(\XX^{(f)}\bigr)^{\!\top}
\Bigl(\ssm\odot\!\!\bigodot_{g\neq f}\!\hat{\hh}^{(g)}\Bigr)
\right],
\label{eq:resonator}
\end{equation}
applied asynchronously, with the factors updated in turn using the most recent estimates of the others. The initial estimates were obtained from the bipolar superposition $\mathrm{sgn}(\sum_{i=1}^{K}\hh^{(f)}_i)$ of each codebook. A trial counts as successful when all $F$ factors are recovered exactly within $T=80$ iterations in any of $N_{\rm restart}=6$ restarts. The first attempt used the superposition initialization; subsequent attempts used randomly signed superpositions $\mathrm{sgn}(\XX^{(f)}\bm{\epsilon})$ with $\bm{\epsilon}\in\{-1,+1\}^{K}$. The success rate was measured over $B$ independent targets per operating point ($B=120$ for $D\leq 65{,}536$, $60$ at $D=131{,}072$, and $40$ above), fitted with a logistic function of $\log_{10}K^{F}$, and $C_{\rm op}$ was defined as the search space at which the fitted success rate reached $0.5$. Because only $N_{\rm restart}=6$ attempts were allowed, the success rate did not reach unity even at negligible load, and the fitted ceiling was treated as a free parameter; averaged over the dimensionalities sampled with four or more codebook sizes, it was $0.83$ for generated HVs and $0.85$ for ideal HVs, and at the two largest dimensionalities, which were sampled with three codebook sizes, the ceiling was held at that average. Since $C_{\rm op}$ depends on $(N_{\rm restart},T)$ as well as on the HVs themselves, the two series are always compared under the same settings.

\subsection*{Optical similarity operation}

The class HVs are binary and are displayed directly on a DMD (Texas Instruments DLP650LNIR; $1280\times800$ mirrors, $10.8\,\mu$m pitch, and a maximum refresh rate of $10.752$ kHz) placed in the speckle field generated by the query (Supplementary Note~1). Two lenses ($f_1=200\,\mathrm{mm}$ and $f_2=50\,\mathrm{mm}$) image the DMD onto the image sensor with a demagnification of $f_2/f_1 = 0.25$. Because the DMD is rotated by $45^{\circ}$ about the optical axis in the present setup, the DMD image overlaps with approximately half of the sensor area. Only this overlapping region is used, resulting in $D=1.64\times10^{5}$. The registration between the sensor-pixel grid and the DMD mirror grid was established by displaying a reference pattern on the DMD and recording it with the sensor; each HV component is written to all mirrors imaged onto its pixel. With the registration fixed, the DMD forms $D$ element-wise products in parallel, and the similarity value is their sum. In the present proof-of-concept, the sum was formed digitally over the registered pixels of the recorded frame; photodetectors placed in the collection path would form it by optical integration, as discussed in Supplementary Note~4.

\subsection*{Iris and MNIST benchmarks}
For the Iris dataset (3 classes, 4 features per sample, 150 samples), the four raw features are mapped into the phase swing of four of the $M=14$ input channels, the remaining channels being held at a fixed bias. Training uses $30$ samples per class (total $90$ samples), and the remaining 60 samples are used for testing, following a stratified random $60/40$ split repeated with three
seeds; the reported accuracy is the mean over the three splits.

For the MNIST dataset, each $28\times28$ image is compressed to a $14$-dimensional feature vector by a convolutional autoencoder with an auxiliary classifier. The encoder comprises three convolutional blocks ($16\to32\to64$ channels) followed by a fully connected layer to $14$ units; the decoder mirrors it with a fully connected layer and three transposed convolutions; the classifier is a fully connected head ($14\to64\to10$). The network is trained on the full labeled MNIST training set with the sum of a reconstruction loss $\mathcal{L}_{\rm rec}=\mathrm{MSE}(\uu_{\rm in},\uu_{\rm rec})$ and a classification loss $\mathcal{L}_{\rm cls}=\mathrm{CrossEntropy}(\yy,\yy_{\rm target})$, so that the $14$ retained features are both reconstructive and discriminative. The $14$ components are mapped directly onto the $M=14$ input channels, one per phase shifter. 

To keep the experimental duration within one speckle correlation time, approximately one hour under the temperature control held to within $\pm 0.5^{\circ}\mathrm{C}$, class HVs are constructed from $10$ images per class ($100$ in total), and a total of $500$ images are used for testing. This restriction is mild here because HDC class prototypes are formed by bundling and converge with few samples per class. In the numerical reference, replacing the $10$ images per class by the full training set changes the accuracy by less than $0.04\%$ on the full test set. The reported accuracy is the mean over five repeated measurements, and the error bar on the experimental result in Fig.~\ref{fig:mnist}(e) is their standard deviation.
 
\paragraph{Numerical simulation.}
The reference simulation of Fig.~\ref{fig:mnist}(c) reproduces the full pipeline through Eq.~\eqref{eq:speckle}, with the transfer matrix $\Am$ modeled, for simplicity, as an i.i.d.\ complex Gaussian random matrix of zero mean and unit variance, neglecting the spatial correlation and inhomogeneous illumination of the measured speckle. The input phase at each phase shifter is modeled as $\phi \propto x^2$, as in the experiment. Query and class HVs are obtained from the simulated intensities, and the optical similarity operation is modeled explicitly as the inner product between the query HV and each class HV. The simulation was repeated on the full test set for five independent random realizations of $\Am$, and the error bar on the numerical result in Fig.~\ref{fig:mnist}(e) is their standard deviation.

\section*{Data availability}
Restrictions apply to the raw image-sensor recordings owing to their volume; the
processed hypervector datasets that support the findings of this study are
available from the corresponding author upon reasonable request.

\section*{Code availability}
The codes used in this study are available from the corresponding author upon reasonable request.

\section*{Acknowledgements}
This work was supported by JSPS KAKENHI (Grant Nos.\ JP22H05198, JP25K22086, JP26K21733) and Japan Science and Technology Agency (JST) CREST (Grant No.\ JPMJCR24R2).

\section*{Author contributions}
S.S.\ conceived the study. N.S.\ and S.S.\ designed the PIC. N.S.\ built the experimental setup, and T.I., N.S., and S.S.\ performed the experiments. T.I.\ performed the numerical simulations and data analysis under the supervision of S.S. T.I.\ and S.S.\ wrote the manuscript. All authors discussed the results and commented on the manuscript.

\section*{Corresponding author}
Correspondence to Satoshi Sunada.

\section*{Competing interests}
The authors declare no competing interests.

\section*{Supplementary information}
Supplementary Information accompanies this paper.

\clearpage
\bibliographystyle{unsrtnat}
\bibliography{samples_ver4}

@article{feldmann2021,
	author = {Feldmann, J. and Youngblood, N. and Karpov, M. and Gehring, H. and Li, X. and Stappers, M. and Le Gallo, M. and Fu, X. and Lukashchuk, A. and Raja, A. S. and Liu, J. and Wright, C. D. and Sebastian, A. and Kippenberg, T. J. and Pernice, W. H. P. and Bhaskaran, H.},
	date = {2021/01/01},
	doi = {10.1038/s41586-020-03070-1},
	id = {Feldmann2021},
	isbn = {1476-4687},
	journal = {Nature},
	number = {7840},
	pages = {52--58},
	title = {Parallel convolutional processing using an integrated photonic tensor core},
	url = {https://doi.org/10.1038/s41586-020-03070-1},
	volume = {589},
	year = {2021}
}

@article{shen2017,
	author = {Shen, Yichen and Harris, Nicholas C. and Skirlo, Scott and Prabhu, Mihika and Baehr-Jones, Tom and Hochberg, Michael and Sun, Xin and Zhao, Shijie and Larochelle, Hugo and Englund, Dirk and Solja{\v c}i{\'c}, Marin},
	date = {2017/07/01},
	doi = {10.1038/nphoton.2017.93},
	id = {Shen2017},
	isbn = {1749-4893},
	journal = {Nature Photonics},
	number = {7},
	pages = {441--446},
	title = {Deep learning with coherent nanophotonic circuits},
	url = {https://doi.org/10.1038/nphoton.2017.93},
	volume = {11},
	year = {2017}
}

@article{McMahon:2023,
  	author = {McMahon, Peter L. },
	date = {2023/12/01},
	doi = {10.1038/s42254-023-00645-5},
	id = {McMahon2023},
	isbn = {2522-5820},
	journal = {Nature Reviews Physics},
	number = {12},
	pages = {717--734},
	title = {The physics of optical computing},
	url = {https://doi.org/10.1038/s42254-023-00645-5},
	volume = {5},
	year = {2023}
}

@article{Miller:2017,
 author={Miller, David A. B.},
  journal={Journal of Lightwave Technology},
  title={{Attojoule Optoelectronics for Low-Energy Information Processing and Communications}},
  year={2017},
  volume={35},
  number={3},
  pages={346-396},
  doi={10.1109/JLT.2017.2647779}
}

@article{Hamerly:2019,
  title = {Large-Scale Optical Neural Networks Based on Photoelectric Multiplication},
  author = {Hamerly, Ryan and Bernstein, Liane and Sludds, Alexander and Solja\ifmmode \check{c}\else \v{c}\fi{}i\ifmmode \acute{c}\else \'{c}\fi{}, Marin and Englund, Dirk},
  journal = {Phys. Rev. X},
  volume = {9},
  issue = {2},
  pages = {021032},
  numpages = {12},
  year = {2019},
  month = {May},
  publisher = {American Physical Society},
  doi = {10.1103/PhysRevX.9.021032},
  url = {https://link.aps.org/doi/10.1103/PhysRevX.9.021032}
}

@article{Chen_Science2025,
author = {Yitong Chen  and Xinyue Sun  and Longtao Tan  and Yizhou Jiang  and Yin Zhou  and Wenjun Zhang  and Guangtao Zhai },
title = {All-optical synthesis chip for large-scale intelligent semantic vision generation},
journal = {Science},
volume = {390},
number = {6779},
pages = {1259-1265},
year = {2025},
URL = {https://www.science.org/doi/abs/10.1126/science.adv7434},
eprint = {https://www.science.org/doi/pdf/10.1126/science.adv7434},
doi = {10.1126/science.adv7434}
}

@article{Meyer:2026,
	author = {Meyer, Lennart and Dijkstra, Jelle and Tebeck, Simon and McRae, Liam and Bahr, Niklas and Steinmeyer, Daniel and Koptyaev, Sergey and Bernasconi, Johana and Pavlov, Nikolay G. and Karpov, Maxim and Jost, John D. and Pernice, Wolfram and Br{\"u}ckerhoff-Pl{\"u}ckelmann, Frank},
	date = {2026/04/09},
	doi = {10.1038/s41467-026-71599-2},
	id = {Meyer2026},
	isbn = {2041-1723},
	journal = {Nature Communications},
	number = {1},
	pages = {3396},
	title = {Deep neural network inference on an integrated, reconfigurable photonic tensor processor},
	url = {https://doi.org/10.1038/s41467-026-71599-2},
	volume = {17},
	year = {2026}
}

@article{Sved:2026,
	author = {Sved, Joel and Song, Shijie and Li, Liwei and Li, George and Meng, Debin and Yi, Xiaoke},
	date = {2026/03/04},
	doi = {10.1038/s41467-026-68648-1},
	id = {Sved2026},
	isbn = {2041-1723},
	journal = {Nature Communications},
	number = {1},
	pages = {1059},
	title = {Inverse-designed nanophotonic neural network accelerators for ultra-compact optical computing},
	url = {https://doi.org/10.1038/s41467-026-68648-1},
	volume = {17},
	year = {2026}
}

@article{Ou:2025,
author = {Shaoyuan Ou  and Kaiwen Xue  and Lian Zhou  and Chun{-}ho Lee  and Alexander Sludds  and Ryan Hamerly  and Ke Zhang  and Hanke Feng  and Yue Yu  and Reshma Kopparapu  and Eric Zhong  and Cheng Wang  and Dirk Englund  and Mengjie Yu  and Zaijun Chen },
title = {Hypermultiplexed integrated photonics–based optical tensor processor},
journal = {Science Advances},
volume = {11},
number = {23},
pages = {eadu0228},
year = {2025},
doi = {10.1126/sciadv.adu0228},
URL = {https://www.science.org/doi/abs/10.1126/sciadv.adu0228},
eprint = {https://www.science.org/doi/pdf/10.1126/sciadv.adu0228}
}

@article{Luan:2026,
	author = {Luan, Chao and Davis III, Ronald and Chen, Zaijun and Englund, Dirk and Hamerly, Ryan},
	date = {2026/01/13},
	doi = {10.1038/s41467-026-68452-x},
	id = {Luan2026},
	isbn = {2041-1723},
	journal = {Nature Communications},
	number = {1},
	pages = {484},
	title = {Single-shot matrix-matrix photonic processor based on spatial-spectral hypermultiplexed parallel diffraction},
	url = {https://doi.org/10.1038/s41467-026-68452-x},
	volume = {17},
	year = {2026}
}

@article{Zhao:2025,
	author = {Zhao, Zhenyu and Pan, Yichen and Xiang, Jinlong and Zhang, Yujia and He, An and Zhao, Yaotian and Chen, Youlve and He, Yu and Fang, Xinyuan and Su, Yikai and Gu, Min and Guo, Xuhan},
	date = {2025/11/21},
	doi = {10.1038/s41467-025-65213-0},
	id = {Zhao2025},
	isbn = {2041-1723},
	journal = {Nature Communications},
	number = {1},
	pages = {10297},
	title = {High computational density nanophotonic media for machine learning inference},
	url = {https://doi.org/10.1038/s41467-025-65213-0},
	volume = {16},
	year = {2025}
}

@article{Bente:2025,
	author = {Bente, Ivonne and Taheriniya, Shabnam and Lenzini, Francesco and Br{\"u}ckerhoff-Pl{\"u}ckelmann, Frank and Kues, Michael and Bhaskaran, Harish and Wright, C. David and Pernice, Wolfram},
	date = {2025/08/01},
	doi = {10.1038/s42254-025-00843-3},
	id = {Bente2025},
	isbn = {2522-5820},
	journal = {Nature Reviews Physics},
	number = {8},
	pages = {439--450},
	title = {The potential of multidimensional photonic computing},
	url = {https://doi.org/10.1038/s42254-025-00843-3},
	volume = {7},
	year = {2025}
}

@article{shastri2021,
	author = {Shastri, Bhavin J. and Tait, Alexander N. and Ferreira de Lima, T. and Pernice, Wolfram H. P. and Bhaskaran, Harish and Wright, C. D. and Prucnal, Paul R.},
	date = {2021/02/01},
	doi = {10.1038/s41566-020-00754-y},
	id = {Shastri2021},
	isbn = {1749-4893},
	journal = {Nature Photonics},
	number = {2},
	pages = {102--114},
	title = {Photonics for artificial intelligence and neuromorphic computing},
	url = {https://doi.org/10.1038/s41566-020-00754-y},
	volume = {15},
	year = {2021}
}

@article{Nahmias:2020,
  author={Nahmias, Mitchell A. and de Lima, Thomas Ferreira and Tait, Alexander N. and Peng, Hsuan-Tung and Shastri, Bhavin J. and Prucnal, Paul R.},
  journal={IEEE Journal of Selected Topics in Quantum Electronics},
  title={{Photonic Multiply-Accumulate Operations for Neural Networks}},
  year={2020},
  volume={26},
  number={1},
  pages={1-18},
  doi={10.1109/JSTQE.2019.2941485}
  }

@article{kanerva2009,
	author = {Kanerva, Pentti},
	date = {2009/06/01},
	doi = {10.1007/s12559-009-9009-8},
	id = {Kanerva2009},
	isbn = {1866-9964},
	journal = {Cognitive Computation},
	number = {2},
	pages = {139--159},
	title = {{Hyperdimensional Computing: An Introduction to Computing in Distributed Representation with High-Dimensional Random Vectors}},
	url = {https://doi.org/10.1007/s12559-009-9009-8},
	volume = {1},
	year = {2009}
}

@book{Plate:2003,
title={{Holographic Reduced Representation: Distributed representation for cognitive structures}},
  author={Plate, Tony A},
  volume={150},
  year={2003},
  publisher={CSLI Publications Stanford},
  isbn      = {978-1-57586-430-3},
}

@inproceedings{Kanerva:1996,
author={Kanerva, Pentti},
editor={von der Malsburg, Christoph
and von Seelen, Werner
and Vorbr{\"u}ggen, Jan C.
and Sendhoff, Bernhard},
title={Binary spatter-coding of ordered K-tuples},
booktitle={Artificial Neural Networks --- ICANN 96},
year={1996},
publisher={Springer Berlin Heidelberg},
address={Berlin, Heidelberg},
pages={869--873},
isbn={978-3-540-68684-2}
}

@article{ge2020,
  author={Ge, Lulu and Parhi, Keshab K.},
  journal={IEEE Circuits and Systems Magazine}, 
  title={{Classification Using Hyperdimensional Computing: A Review}}, 
  year={2020},
  volume={20},
  number={2},
  pages={30-47},
  doi={10.1109/MCAS.2020.2988388}}

@inproceedings{rahimi2016,
author = {Rahimi, Abbas and Kanerva, Pentti and Rabaey, Jan M.},
title = {{A Robust and Energy-Efficient Classifier Using Brain-Inspired Hyperdimensional Computing}},
year = {2016},
isbn = {9781450341851},
publisher = {Association for Computing Machinery},
address = {New York, NY, USA},
url = {https://doi.org/10.1145/2934583.2934624},
doi = {10.1145/2934583.2934624},
booktitle = {Proceedings of the 2016 International Symposium on Low Power Electronics and Design},
pages = {64–69},
numpages = {6},
location = {San Francisco Airport, CA, USA},
series = {ISLPED '16}
}

@article{thomas2021,
author = {Thomas, Anthony and Dasgupta, Sanjoy and Rosing, Tajana},
title = {{A Theoretical Perspective on Hyperdimensional Computing}},
year = {2022},
issue_date = {Jan 2022},
publisher = {AI Access Foundation},
address = {El Segundo, CA, USA},
volume = {72},
issn = {1076-9757},
url = {https://doi.org/10.1613/jair.1.12664},
doi = {10.1613/jair.1.12664},
journal = {J. Artif. Int. Res.},
pages = {215-249},
numpages = {35}
}

@article{Kleyko:2022a,
author = {Kleyko, Denis and Rachkovskij, Dmitri A. and Osipov, Evgeny and Rahimi, Abbas},
title = {{A Survey on Hyperdimensional Computing aka Vector Symbolic Architectures, Part I: Models and Data Transformations}},
year = {2022},
issue_date = {June 2023},
publisher = {Association for Computing Machinery},
address = {New York, NY, USA},
volume = {55},
number = {6},
issn = {0360-0300},
url = {https://doi.org/10.1145/3538531},
doi = {10.1145/3538531},
journal = {ACM Comput. Surv.},
month = dec,
articleno = {130},
numpages = {40}
}

@INPROCEEDINGS{HernandezCano:2021,
  author={Hern{\'a}ndez-Cano, Alejandro and Matsumoto, Namiko and Ping, Eric and Imani, Mohsen},
  booktitle={2021 Design, Automation \& Test in Europe Conference \& Exhibition (DATE)}, 
  title={{OnlineHD: Robust, Efficient, and Single-Pass Online Learning Using Hyperdimensional System}}, 
  year={2021},
  volume={},
  number={},
  pages={56-61},
  doi={10.23919/DATE51398.2021.9474107}}

@article{Hersche:2023,
	author = {Hersche, Michael and Zeqiri, Mustafa and Benini, Luca and Sebastian, Abu and Rahimi, Abbas},
	date = {2023/04/01},
	doi = {10.1038/s42256-023-00630-8},
	id = {Hersche2023},
	isbn = {2522-5839},
	journal = {Nature Machine Intelligence},
	number = {4},
	pages = {363--375},
	title = {{A neuro-vector-symbolic architecture for solving Raven's progressive matrices}},
	url = {https://doi.org/10.1038/s42256-023-00630-8},
	volume = {5},
	year = {2023}
}

@article{Karunaratne:2021,
	author = {Karunaratne, Geethan and Schmuck, Manuel and Le Gallo, Manuel and Cherubini, Giovanni and Benini, Luca and Sebastian, Abu and Rahimi, Abbas},
	date = {2021/04/29},
	doi = {10.1038/s41467-021-22364-0},
	id = {Karunaratne2021},
	isbn = {2041-1723},
	journal = {Nature Communications},
	number = {1},
	pages = {2468},
	title = {Robust high-dimensional memory-augmented neural networks},
	url = {https://doi.org/10.1038/s41467-021-22364-0},
	volume = {12},
	year = {2021}
}

@article{Langenegger:2023,
	author = {Langenegger, Jovin and Karunaratne, Geethan and Hersche, Michael and Benini, Luca and Sebastian, Abu and Rahimi, Abbas},
	date = {2023/05/01},
	doi = {10.1038/s41565-023-01357-8},
	id = {Langenegger2023},
	isbn = {1748-3395},
	journal = {Nature Nanotechnology},
	number = {5},
	pages = {479--485},
	title = {In-memory factorization of holographic perceptual representations},
	url = {https://doi.org/10.1038/s41565-023-01357-8},
	volume = {18},
	year = {2023}
}

@INPROCEEDINGS{Rahimi:2016b,
  author={Rahimi, Abbas and Benatti, Simone and Kanerva, Pentti and Benini, Luca and Rabaey, Jan M.},
  booktitle={2016 IEEE International Conference on Rebooting Computing (ICRC)}, 
  title={{Hyperdimensional biosignal processing: A case study for EMG-based hand gesture recognition}}, 
  year={2016},
  volume={},
  number={},
  pages={1-8},
  doi={10.1109/ICRC.2016.7738683}}

@INPROCEEDINGS{Imani:HDNA:2018,
  author={Imani, Mohsen and Nassar, Tarek and Rahimi, Abbas and Rosing, Tajana},
  booktitle={2018 IEEE EMBS International Conference on Biomedical and Health Informatics (BHI)}, 
  title={{HDNA: Energy-efficient DNA sequencing using hyperdimensional computing}}, 
  year={2018},
  volume={},
  number={},
  pages={271-274},
  doi={10.1109/BHI.2018.8333421}}

@inproceedings{Zhuowen_BioHD:2022,
author = {Zou, Zhuowen and Chen, Hanning and Poduval, Prathyush and Kim, Yeseong and Imani, Mahdi and Sadredini, Elaheh and Cammarota, Rosario and Imani, Mohsen},
title = {{BioHD: an efficient genome sequence search platform using HyperDimensional memorization}},
year = {2022},
isbn = {9781450386104},
publisher = {Association for Computing Machinery},
address = {New York, NY, USA},
url = {https://doi.org/10.1145/3470496.3527422},
doi = {10.1145/3470496.3527422},
booktitle = {Proceedings of the 49th Annual International Symposium on Computer Architecture},
pages = {656–669},
numpages = {14},
location = {New York, New York},
series = {ISCA '22}
}

@article{Mitrokhin:2019,
author = {A. Mitrokhin  and P. Sutor  and C. Fermüller  and Y. Aloimonos },
title = {Learning sensorimotor control with neuromorphic sensors: Toward hyperdimensional active perception},
journal = {Science Robotics},
volume = {4},
number = {30},
pages = {eaaw6736},
year = {2019},
doi = {10.1126/scirobotics.aaw6736},
URL = {https://www.science.org/doi/abs/10.1126/scirobotics.aaw6736},
eprint = {https://www.science.org/doi/pdf/10.1126/scirobotics.aaw6736}
}

@article{Yue:2026,
	author = {Yue, Yuanli and Gouda, Muhammed and Sunada, Satoshi and Bienstman, Peter},
	date = {2026/03/25},
	doi = {10.1038/s41598-026-44705-z},
	id = {Yue2026},
	isbn = {2045-2322},
	journal = {Scientific Reports},
	number = {1},
	pages = {14900},
	title = {Hyper-dimensional computing for enhanced label-free particle analysis in a flow-based optical detection system},
	url = {https://doi.org/10.1038/s41598-026-44705-z},
	volume = {16},
	year = {2026}
}

@article{Kitagawa:24,
author = {Kei Kitagawa and Kohei Tsuji and Koyo Sagehashi and Tomoaki Niiyama and Satoshi Sunada},
journal = {Opt. Express},
number = {3},
pages = {3209--3220},
publisher = {Optica Publishing Group},
title = {Optical hyperdimensional soft sensing: speckle-based touch interface and tactile sensor},
volume = {32},
month = {Jan},
year = {2024},
url = {https://opg.optica.org/oe/abstract.cfm?URI=oe-32-3-3209},
doi = {10.1364/OE.513802}
}

@article{Kleyko:2023,
author = {Kleyko, Denis and Rachkovskij, Dmitri and Osipov, Evgeny and Rahimi, Abbas},
title = {{A Survey on Hyperdimensional Computing aka Vector Symbolic Architectures, Part II: Applications, Cognitive Models, and Challenges}},
year = {2023},
issue_date = {September 2023},
publisher = {Association for Computing Machinery},
address = {New York, NY, USA},
volume = {55},
number = {9},
issn = {0360-0300},
url = {https://doi.org/10.1145/3558000},
doi = {10.1145/3558000},
journal = {ACM Comput. Surv.},
month = jan,
articleno = {175},
numpages = {52}
}

@article{Schlegel2022,
	author = {Schlegel, Kenny and Neubert, Peer and Protzel, Peter},
	date = {2022/08/01},
	doi = {10.1007/s10462-021-10110-3},
	id = {Schlegel2022},
	isbn = {1573-7462},
	journal = {Artificial Intelligence Review},
	number = {6},
	pages = {4523--4555},
	title = {A comparison of vector symbolic architectures},
	url = {https://doi.org/10.1007/s10462-021-10110-3},
	volume = {55},
	year = {2022}
}

@inproceedings{Arockiaraj:2026,
author = {Arockiaraj, Jebacyril and Parikh, Dhruv and Prasanna, Viktor},
title = {{NysX: An Accurate and Energy-Efficient FPGA Accelerator for Hyperdimensional Graph Classification at the Edge}},
year = {2026},
isbn = {9798400720796},
publisher = {Association for Computing Machinery},
address = {New York, NY, USA},
url = {https://doi.org/10.1145/3748173.3779549},
doi = {10.1145/3748173.3779549},
booktitle = {Proceedings of the 2026 ACM/SIGDA International Symposium on Field Programmable Gate Arrays},
pages = {140},
numpages = {1},
location = {USA},
series = {FPGA '26}
}

@ARTICLE{Kleyko:2022b,
  author={Kleyko, Denis and Davies, Mike and Frady, Edward Paxon and Kanerva, Pentti and Kent, Spencer J. and Olshausen, Bruno A. and Osipov, Evgeny and Rabaey, Jan M. and Rachkovskij, Dmitri A. and Rahimi, Abbas and Sommer, Friedrich T.},
  journal={Proceedings of the IEEE}, 
  title={{Vector Symbolic Architectures as a Computing Framework for Emerging Hardware}}, 
  year={2022},
  volume={110},
  number={10},
  pages={1538-1571},
  doi={10.1109/JPROC.2022.3209104}}

@article{karunaratne2020,
	author = {Karunaratne, Geethan and Le Gallo, Manuel and Cherubini, Giovanni and Benini, Luca and Rahimi, Abbas and Sebastian, Abu},
	date = {2020/06/01},
	doi = {10.1038/s41928-020-0410-3},
	id = {Karunaratne2020},
	isbn = {2520-1131},
	journal = {Nature Electronics},
	number = {6},
	pages = {327--337},
	title = {In-memory hyperdimensional computing},
	url = {https://doi.org/10.1038/s41928-020-0410-3},
	volume = {3},
	year = {2020}}

@ARTICLE{Wu:2018,
  author={Wu, Tony F. and Li, Haitong and Huang, Ping-Chen and Rahimi, Abbas and Hills, Gage and Hodson, Bryce and Hwang, William and Rabaey, Jan M. and Wong, H.-S. Philip and Shulaker, Max M. and Mitra, Subhasish},
  journal={IEEE Journal of Solid-State Circuits}, 
  title={{Hyperdimensional Computing Exploiting Carbon Nanotube FETs, Resistive RAM, and Their Monolithic 3D Integration}}, 
  year={2018},
  volume={53},
  number={11},
  pages={3183-3196},
  doi={10.1109/JSSC.2018.2870560}}

@article{Fayza:2026,
author = {Fayza, Farbin and Demirkiran, Cansu and Yang, Guowei and Chen, Hanning and Liu, Che-Kai and Mohan, Avi and Errahmouni, Hamza and Yun, Sanggeon and Imani, Mohsen and Zhang, David and Bunandar, Darius and Joshi, Ajay},
title = {{PhotoHDC: An Electro-Photonic Accelerator for Hyperdimensional Computing}},
year = {2026},
issue_date = {July 2026},
publisher = {Association for Computing Machinery},
address = {New York, NY, USA},
volume = {22},
number = {3},
issn = {1550-4832},
url = {https://doi.org/10.1145/3802583},
doi = {10.1145/3802583},
journal = {J. Emerg. Technol. Comput. Syst.},
month = jun,
articleno = {16},
numpages = {25}
}

@ARTICLE{Najafi:2025,
  author={Najafi, Deniz and Barkam, Hamza Errahmouni and Morsali, Mehrdad and Jeong, SungHeon and Das, Tamoghno and Roohi, Arman and Nikdast, Mahdi and Imani, Mohsen and Angizi, Shaahin},
  journal={IEEE Transactions on Circuits and Systems for Artificial Intelligence}, 
  title={{Neuro-Photonix: Enabling Near-Sensor Neuro-Symbolic AI Computing on Silicon Photonics Substrate}}, 
  year={2025},
  volume={2},
  number={2},
  pages={101-113},
  doi={10.1109/TCASAI.2025.3537968}}

@inproceedings{LiuMa:2025,
author = {Liu, Jiaqi and Ma, Yiwen},
title = {{OpticalHDC: Ultra-fast Photonic Hyperdimensional Computing Accelerator}},
year = {2025},
isbn = {9798400706356},
publisher = {Association for Computing Machinery},
address = {New York, NY, USA},
url = {https://doi.org/10.1145/3658617.3697709},
doi = {10.1145/3658617.3697709},
booktitle = {Proceedings of the 30th Asia and South Pacific Design Automation Conference},
pages = {748–753},
numpages = {6},
location = {Tokyo, Japan},
series = {ASPDAC '25}
}

@article{Schmuck:2019,
author = {Schmuck, Manuel and Benini, Luca and Rahimi, Abbas},
title = {{Hardware Optimizations of Dense Binary Hyperdimensional Computing: Rematerialization of Hypervectors, Binarized Bundling, and Combinational Associative Memory}},
year = {2019},
issue_date = {October 2019},
publisher = {Association for Computing Machinery},
address = {New York, NY, USA},
volume = {15},
number = {4},
issn = {1550-4832},
url = {https://doi.org/10.1145/3314326},
doi = {10.1145/3314326},
journal = {J. Emerg. Technol. Comput. Syst.},
month = oct,
articleno = {32},
numpages = {25}
}

@ARTICLE{Kleyko:2022c,
  author={Kleyko, Denis and Frady, Edward Paxon and Sommer, Friedrich T.},
  journal={IEEE Transactions on Neural Networks and Learning Systems}, 
  title={{Cellular Automata Can Reduce Memory Requirements of Collective-State Computing}}, 
  year={2022},
  volume={33},
  number={6},
  pages={2701-2713},
  doi={10.1109/TNNLS.2021.3119543}}

@misc{Thomas_streaming2023,
      title={{Streaming Encoding Algorithms for Scalable Hyperdimensional Computing}}, 
      author={Anthony Thomas and Behnam Khaleghi and Gopi Krishna Jha and Sanjoy Dasgupta and Nageen Himayat and Ravi Iyer and Nilesh Jain and Tajana Rosing},
      year={2023},
      eprint={2209.09868},
      archivePrefix={arXiv},
      primaryClass={cs.LG},
      url={https://arxiv.org/abs/2209.09868}, 
}

@article{Tang:2026,
	author = {Tang, Haoyu and Kan, Yirong and Wu, Man and Zhang, Renyuan and Nakashima, Yasuhiko},
	date = {2026/02/09},
	doi = {10.1007/s11227-026-08276-0},
	id = {Tang2026},
	isbn = {1573-0484},
	journal = {The Journal of Supercomputing},
	number = {3},
	pages = {141},
	title = {{DystoHD: an area-efficient hyperdimensional computing system with dynamic hypervector generation for memory-constrained devices}},
	url = {https://doi.org/10.1007/s11227-026-08276-0},
	volume = {82},
	year = {2026}}

@ARTICLE{Saeidi:2026,
  author={Saeidi, Akbar and Verreault, Antoine and Ahmed, Hassaan and Khaliq, Mehboob and Brun, Damien and Robichaud, Alexandre},
  journal={IEEE Access}, 
  title={{High-Efficiency Hardware Encoding for Hyperdimensional Computing on FPGA}}, 
  year={2026},
  volume={14},
  number={},
  pages={102811-102826},
  doi={10.1109/ACCESS.2026.3711522}}

@misc{Aygun2023_learning,
      title={{Learning from Hypervectors: A Survey on Hypervector Encoding}}, 
      author={Sercan Aygun and Mehran Shoushtari Moghadam and M. Hassan Najafi and Mohsen Imani},
      year={2023},
      eprint={2308.00685},
      archivePrefix={arXiv},
      primaryClass={cs.LG},
      url={https://arxiv.org/abs/2308.00685}, 
}

@inproceedings{saade2016,
author = {Saade, A. and Caltagirone, F. and Carron, I. and Daudet, L. and Dr{\'e}meau, A. and Gigan, S. and Krzakala, F.},
title = {Random projections through multiple optical scattering: Approximating Kernels at the speed of light},
year = {2016},
publisher = {IEEE Press},
url = {https://doi.org/10.1109/ICASSP.2016.7472872},
doi = {10.1109/ICASSP.2016.7472872},
booktitle = {2016 IEEE International Conference on Acoustics, Speech and Signal Processing (ICASSP)},
pages = {6215–6219},
numpages = {5},
location = {Shanghai, China}
}

@article{rafayelyan2020,
  title = {{Large-Scale Optical Reservoir Computing for Spatiotemporal Chaotic Systems Prediction}},
  author = {Rafayelyan, Mushegh and Dong, Jonathan and Tan, Yongqi and Krzakala, Florent and Gigan, Sylvain},
  journal = {Phys. Rev. X},
  volume = {10},
  issue = {4},
  pages = {041037},
  numpages = {11},
  year = {2020},
  month = {Nov},
  publisher = {American Physical Society},
  doi = {10.1103/PhysRevX.10.041037},
  url = {https://link.aps.org/doi/10.1103/PhysRevX.10.041037}
}

@article{Sunada:20,
author = {Satoshi Sunada and Kazutaka Kanno and Atsushi Uchida},
journal = {Opt. Express},
number = {21},
pages = {30349--30361},
publisher = {Optica Publishing Group},
title = {Using multidimensional speckle dynamics for high-speed, large-scale, parallel photonic computing},
volume = {28},
month = {Oct},
year = {2020},
url = {https://opg.optica.org/oe/abstract.cfm?URI=oe-28-21-30349},
doi = {10.1364/OE.399495},
}

@article{Wang_Light2025,
	author = {Wang, Hao and Hu, Jianqi and Baek, YoonSeok and Tsuchiyama, Kohei and Joly, Malo and Liu, Qiang and Gigan, Sylvain},
	date = {2025/07/21},
	doi = {10.1038/s41377-025-01927-6},
	id = {Wang2025},
	isbn = {2047-7538},
	journal = {Light: Science \& Applications},
	number = {1},
	pages = {245},
	title = {Optical next generation reservoir computing},
	url = {https://doi.org/10.1038/s41377-025-01927-6},
	volume = {14},
	year = {2025}
	}

@book{Goodman:2007,
  title={Speckle phenomena in optics: theory and applications},
  author={Goodman, Joseph W},
  year={2007},
  publisher={Roberts and company Publishers}
}

@incollection{Uchida_inbook:2012,
author={Uchida, Atsushi},
publisher = {John Wiley \& Sons, Ltd},
isbn = {9783527640331},
title = {{Random Number Generation with Chaotic Lasers}},
booktitle = {{Optical Communication with Chaotic Lasers}},
chapter = {10},
pages = {445--509},
doi = {https://doi.org/10.1002/9783527640331.ch10},
url = {https://onlinelibrary.wiley.com/doi/abs/10.1002/9783527640331.ch10},
eprint = {https://onlinelibrary.wiley.com/doi/pdf/10.1002/9783527640331.ch10},
year = {2012}
}

@article{Frady:2020,
    author = {Frady, E. Paxon and Kent, Spencer J. and Olshausen, Bruno A. and Sommer, Friedrich T.},
    title = {{Resonator Networks, 1: An Efficient Solution for Factoring High-Dimensional, Distributed Representations of Data Structures}},
    journal = {Neural Computation},
    volume = {32},
    number = {12},
    pages = {2311-2331},
    year = {2020},
    month = {12},
    issn = {0899-7667},
    doi = {10.1162/neco_a_01331},
    url = {https://doi.org/10.1162/neco_a_01331},
    eprint = {https://direct.mit.edu/neco/article-pdf/32/12/2311/1865557/neco_a_01331.pdf}
}

@article{Kent:2020,
    author = {Kent, Spencer J. and Frady, E. Paxon and Sommer, Friedrich T. and Olshausen, Bruno A.},
    title = {{Resonator Networks, 2: Factorization Performance and Capacity Compared to Optimization-Based Methods}},
    journal = {Neural Computation},
    volume = {32},
    number = {12},
    pages = {2332-2388},
    year = {2020},
    month = {12},
    issn = {0899-7667},
    doi = {10.1162/neco_a_01329},
    url = {https://doi.org/10.1162/neco_a_01329},
    eprint = {https://direct.mit.edu/neco/article-pdf/32/12/2332/1865599/neco_a_01329.pdf}
}

@ARTICLE{LeCun:1998,
  author={Lecun, Y. and Bottou, L. and Bengio, Y. and Haffner, P.},
  journal={Proceedings of the IEEE}, 
  title={Gradient-based learning applied to document recognition}, 
  year={1998},
  volume={86},
  number={11},
  pages={2278-2324},
  doi={10.1109/5.726791}}

@Article{W-C_Su:2012,
AUTHOR = {Su, Wei-Chia and Sun, Ching-Cherng},
TITLE = {{Review of Random Phase Encoding in Volume Holographic Storage}},
JOURNAL = {Materials},
VOLUME = {5},
YEAR = {2012},
NUMBER = {9},
PAGES = {1635--1653},
URL = {https://www.mdpi.com/1996-1944/5/9/1635},
ISSN = {1996-1944},
DOI = {10.3390/ma5091635}
}

@article{Butow:2024,
	author = {B{\"u}tow, Johannes and Eismann, J{\"o}rg S. and Sharma, Varun and Brandm{\"u}ller, Dorian and Banzer, Peter},
	date = {2024/03/01},
	doi = {10.1038/s41566-023-01354-2},
	id = {B{\"u}tow2024},
	isbn = {1749-4893},
	journal = {Nature Photonics},
	number = {3},
	pages = {243--249},
	title = {Generating free-space structured light with programmable integrated photonics},
	url = {https://doi.org/10.1038/s41566-023-01354-2},
	volume = {18},
	year = {2024}}

@article{Hu:2023,
    author = {Hu, Gaolei and Zhong, Keyi and Qin, Yue and Tsang, Hon Ki},
    title = {Silicon photonic integrated circuit for high-resolution multimode fiber imaging system},
    journal = {APL Photonics},
    volume = {8},
    number = {4},
    pages = {046104},
    year = {2023},
    month = {04},
    issn = {2378-0967},
    doi = {10.1063/5.0137688},
    url = {https://doi.org/10.1063/5.0137688},
    eprint = {https://pubs.aip.org/aip/app/article-pdf/doi/10.1063/5.0137688/16821938/046104_1_5.0137688.pdf},
}

@article{Wang_NP:2023,
	author = {Wang, Tianyu and Sohoni, Mandar M. and Wright, Logan G. and Stein, Martin M. and Ma, Shi-Yuan and Onodera, Tatsuhiro and Anderson, Maxwell G. and McMahon, Peter L.},
	date = {2023/05/01},
	doi = {10.1038/s41566-023-01170-8},
	id = {Wang2023},
	isbn = {1749-4893},
	journal = {Nature Photonics},
	number = {5},
	pages = {408--415},
	title = {Image sensing with multilayer nonlinear optical neural networks},
	url = {https://doi.org/10.1038/s41566-023-01170-8},
	volume = {17},
	year = {2023}}

@article{DAMMANN1971312,
title = {High-efficiency in-line multiple imaging by means of multiple phase holograms},
journal = {Optics Communications},
volume = {3},
number = {5},
pages = {312-315},
year = {1971},
issn = {0030-4018},
doi = {https://doi.org/10.1016/0030-4018(71)90095-2},
url = {https://www.sciencedirect.com/science/article/pii/0030401871900952},
author = {H. Dammann and K. Görtler}
}

@article{Ma_Dammanngrating:23,
author = {Guoqing Ma and Junjie Yu and Rongwei Zhu and Fenglu Zheng and Changhe Zhou and Guohai Situ},
journal = {Opt. Lett.},
number = {9},
pages = {2301--2304},
publisher = {Optica Publishing Group},
title = {Dammann gratings-based truly parallel optical matrix multiplication accelerator},
volume = {48},
month = {May},
year = {2023},
url = {https://opg.optica.org/ol/abstract.cfm?URI=ol-48-9-2301},
doi = {10.1364/OL.487676},
}

@article{Akiyama:2012,
author = {Suguru Akiyama and Takeshi Baba and Masahiko Imai and Takeshi Akagawa and Masashi Takahashi and Naoki Hirayama and Hiroyuki Takahashi and Yoshiji Noguchi and Hideaki Okayama and Tsuyoshi Horikawa and Tatsuya Usuki},
journal = {Opt. Express},
number = {3},
pages = {2911--2923},
publisher = {Optica Publishing Group},
title = {{12.5-Gb/s operation with 0.29-V{\textperiodcentered}cm V$\pi$L using silicon Mach-Zehnder modulator based-on forward-biased pin diode}},
volume = {20},
month = {Jan},
year = {2012},
url = {https://opg.optica.org/oe/abstract.cfm?URI=oe-20-3-2911},
doi = {10.1364/OE.20.002911},
}

@INPROCEEDINGS{Huang_DAC:2020,
  author={Huang, Hung-Yi and Chen, Xin-Yu and Kuo, Tai-Haur},
  booktitle={2020 IEEE Symposium on VLSI Circuits}, 
  title={{A 177mW 10GS/s NRZ DAC with Switching-Glitch Compensation Achieving $>$ 64dBc SFDR and $<$ $-$77dBc IM3}}, 
  year={2020},
  volume={},
  number={},
  pages={1-2},
  doi={10.1109/VLSICircuits18222.2020.9162931}}

@INPROCEEDINGS{Guo_ADC:2019,
  author={Guo, Mingqiang and Mao, Jiaji and Sin, Sai-Weng and Wei, Hegong and Martins, R. P.},
  booktitle={2019 Symposium on VLSI Circuits}, 
  title={{A 29mW 5GS/s Time-interleaved SAR ADC achieving 48.5dB SNDR With Fully-Digital Timing-Skew Calibration Based on Digital-Mixing}}, 
  year={2019},
  volume={},
  number={},
  pages={C76-C77},
  doi={10.23919/VLSIC.2019.8778077}}

@article{Wang_TIA:2013,
author = {Wang, Xiaoxia and Wang, Zhigong},
title = {Low-power, low-area preamplifier for ultra high-speed multi-channel optical communication system},
journal = {International Journal of Circuit Theory and Applications},
volume = {41},
number = {2},
pages = {186-204},
doi = {https://doi.org/10.1002/cta.794},
url = {https://onlinelibrary.wiley.com/doi/abs/10.1002/cta.794},
eprint = {https://onlinelibrary.wiley.com/doi/pdf/10.1002/cta.794},
year = {2013}
}

@misc{TI_DLP650LNIR,
  author       = {{Texas Instruments}},
  title        = {{DLP650LNIR} 0.65 {WXGA} {NIR} {DMD} data sheet},
  howpublished = {\url{https://www.ti.com/product/DLP650LNIR}},
  year         = {2026},
  note         = {Accessed: 2026-09-16}
}

@misc{QHY991,
  author       = {{QHYCCD}},
  title        = {{QHY990/QHY991/QHY992 SWIR} scientific cameras: specifications},
  howpublished = {\url{https://www.qhyccd.com/scientific-camera-qhy990_qhy991_qhy992/}},
  year         = {2026},
  note         = {Accessed: 2026-09-16}
}

@misc{TI_DLP9500,
  author       = {{Texas Instruments}},
  title        = {{DLP9500} data sheet, product information and support},
  howpublished = {\url{https://www.ti.com/product/DLP9500}},
  year         = {2026},
  note         = {Accessed: 2026-08-16}
}

@INPROCEEDINGS{Peng_NeuroSim:2019,
  author={Peng, Xiaochen and Huang, Shanshi and Luo, Yandong and Sun, Xiaoyu and Yu, Shimeng},
  booktitle={2019 IEEE International Electron Devices Meeting (IEDM)}, 
  title={{DNN+NeuroSim: An End-to-End Benchmarking Framework for Compute-in-Memory Accelerators with Versatile Device Technologies}}, 
  year={2019},
  volume={},
  number={},
  pages={32.5.1-32.5.4},
  doi={10.1109/IEDM19573.2019.8993491}}

@misc{Fisher_Iris_UCI,
  author = {Fisher, R.~A.},
  title = {{Iris} [Dataset]},
  year = {1936},
  howpublished = {UCI Machine Learning Repository},
  doi = {10.24432/C56C76},
  url = {https://doi.org/10.24432/C56C76}
}

\end{document}


\begin{center}
{\Large\bfseries Supplementary Information}\\[6pt]
{\large Hypervectors from Optical Disorder: Programmable Encoding and Optical Inference for Hyperdimensional Computing}\\[6pt]
Takuya Iwata, Namthip Srisuthep, and Satoshi Sunada
\end{center}

\vspace{1em}

\section{Details on the experimental setup}
\label{sec:expset}

\begin{figure}[H]
  \centering
  \includegraphics[width=0.7\textwidth]{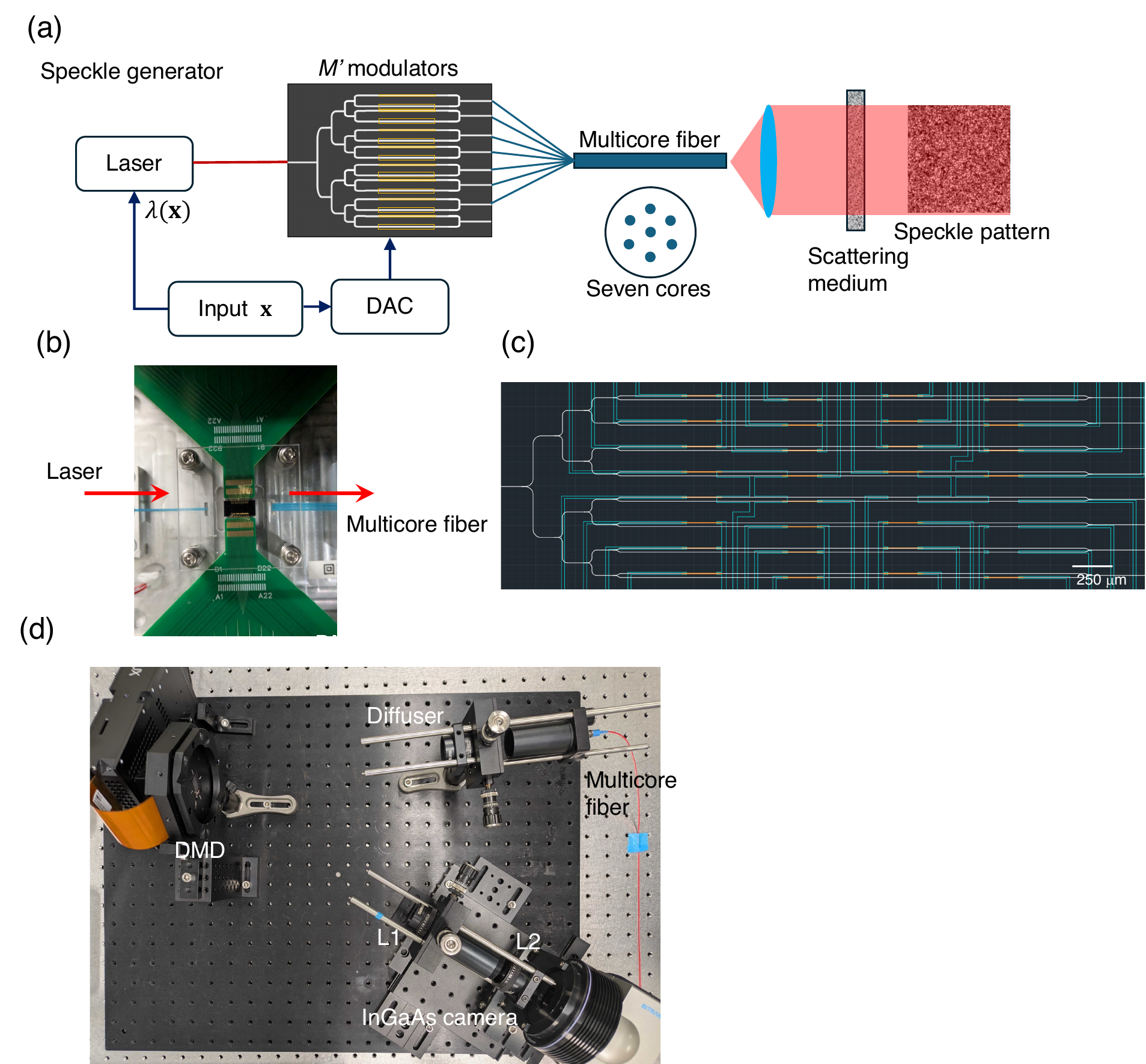}
 \caption{\textbf{Details of the experimental setup.} (a) Schematic of the speckle generator. The input $\xx$ drives the $M$ phase shifters of the MZM array through a DAC and, optionally, controls the laser wavelength $\lambda(\xx)$. The modulated optical fields are relayed through the seven cores of a multicore fiber, collimated, and superposed on the diffuser to form the speckle pattern. (b) Photograph of the PIC, edge-coupled to the laser input on one side and to the multicore fiber on the other. (c) Layout of the MZM array inside the PIC. (d) Photograph of the free-space section for optical similarity-based inference, comprising the diffuser, the DMD, collection lenses $L_1$ and $L_2$ with focal lengths of $200\,\mathrm{mm}$ and $50\,\mathrm{mm}$, respectively, and the InGaAs camera.}
  \label{fig:expset}
\end{figure}

Figure~\ref{fig:expset} shows the experimental setup described in Methods. The speckle generator consists primarily of a laser source, an on-chip MZM array, a seven-core multicore fiber, and a diffuser (scattering medium) [Fig.~\ref{fig:expset}(a)]. The multicore fiber couples the outputs of the MZM array into the free-space optical section. The top view and detailed layout of the on-chip MZM array are shown in Fig.~\ref{fig:expset}(b) and (c), respectively. The free-space setup shown in Fig.~\ref{fig:expset}(d) is used for optical similarity-based inference.

\section{Estimation of $N_{\rm sp}$ and noise influence}\label{sec:nsp}
\subsection*{Spatial correlation and estimation of $N_{\rm sp}$}
\begin{figure}[H]
  \centering
  \includegraphics[width=0.5\textwidth]{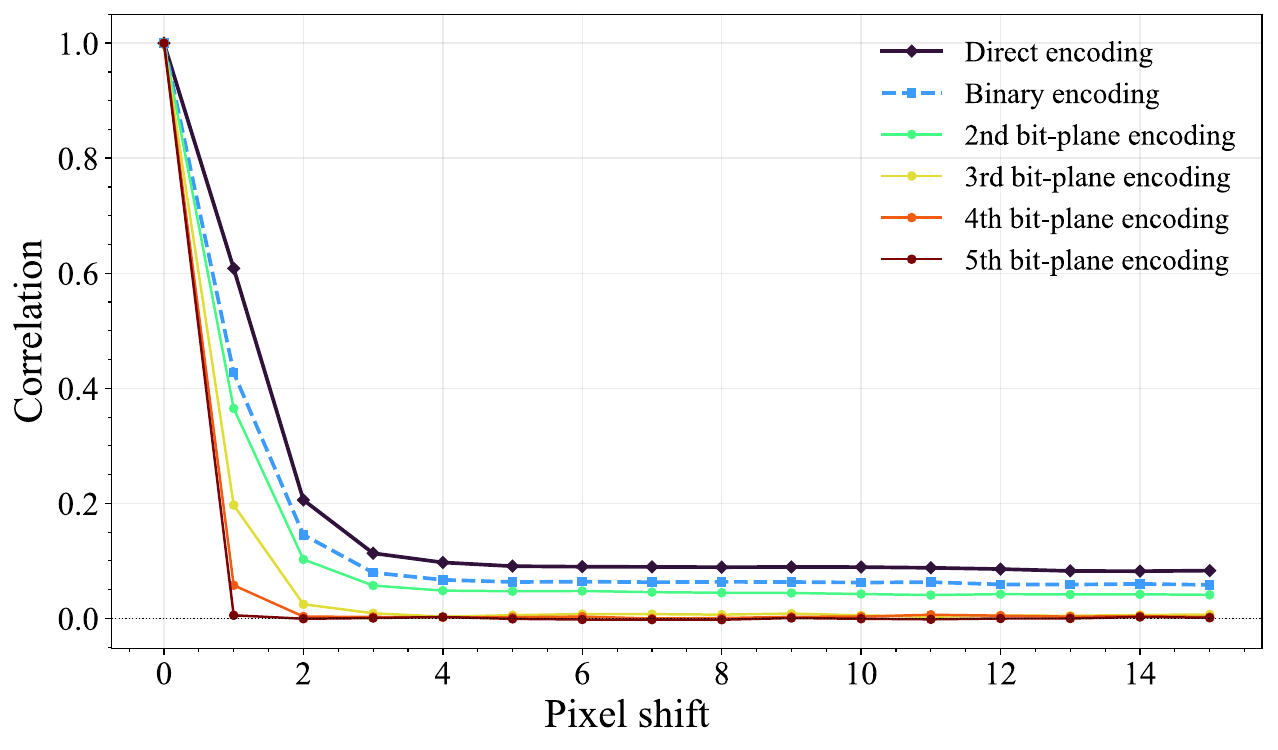}
\caption{\textbf{Spatial autocorrelation of the generated HVs.} Spatial autocorrelation versus pixel shift for the zero-mean and unit-variance normalized raw speckle pattern (direct encoding), the binary HV, and the $k$-th bit-plane encoded HVs with $k=2$--$5$.}
  \label{fig:si_corr}
\end{figure}

Figure~\ref{fig:si_corr} shows the spatial autocorrelation of the raw speckle
pattern (direct encoding), the corresponding binary HV, and the $k$-th bit-plane HVs. The floor at small nonzero correlation for the direct, binary and $k=2$
encodings is attributed to the slowly varying illumination envelope, and vanishes for $k\geq3$. 

The number of independent grains can be roughly estimated as $N_{\rm sp}\simeq D_{\max}/n_{\rm corr}^{2}$ with $D_{\max}=327{,}680$, where $n_{\rm corr}$ denotes the correlation length defined as the smallest pixel shift at which the autocorrelation reaches its asymptotic floor. Direct encoding gives $n_{\rm corr}=4$ and $N_{\rm sp} \simeq 2\times10^{4}$. 

For the $k$-th bit-plane encoding, $n_{\rm corr}$ decreases with increasing $k$ and reaches 1 at $k=5$, indicating that the bit-plane extraction deterministically expands the randomness of the original field. Thus, the physical information content is bounded by the grain count rather than the HV dimensionality, and the effective HDC dimensionality is assessed empirically from the similarity and capacity measurements in the main text.

\subsection*{Readout noise in bit planes}
\begin{figure}[H]
  \centering
  \includegraphics[width=0.5\textwidth]{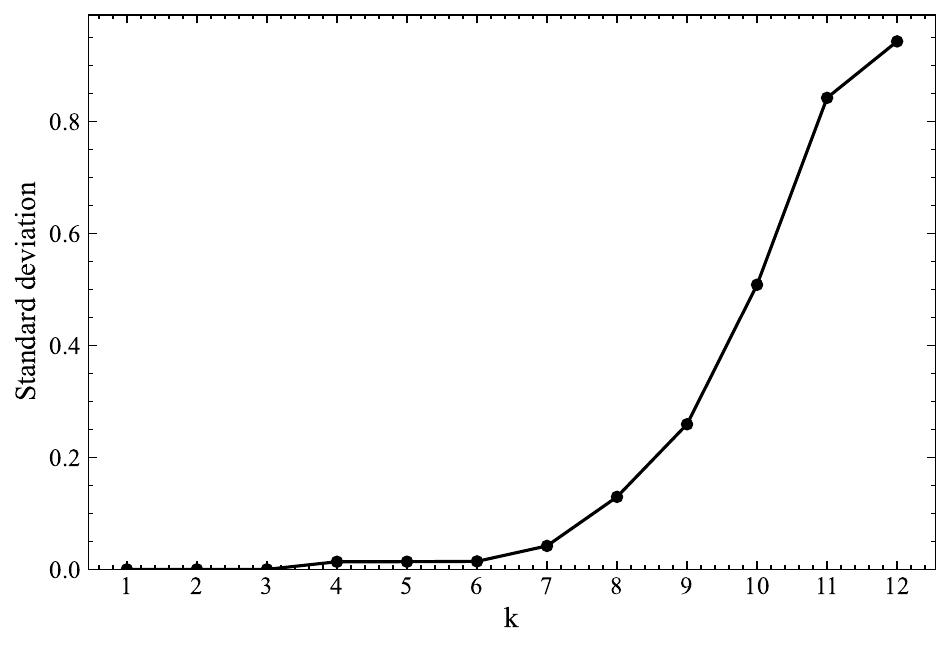}
\caption{\textbf{Standard deviation of bipolar components induced by readout noise for each bit plane $k$.}}
  \label{fig:si_noise}
\end{figure}

To quantify the effect of camera readout noise on the $k$-th bit plane for $Q=12$, we recorded 10 images and calculated the standard deviation of each bipolar component across the frames, averaged over all components. The results are shown in Fig.~\ref{fig:si_noise}. The standard deviation remains approximately 0.01 for $k\leq6$.

\section{Number of quasi-orthogonal HVs the generator can supply}
\label{sec:nhv}
Here, we roughly estimate the number $N_{\rm HV}$ of quasi-orthogonal HVs that the generator can supply using two addressing channels: the input phase vector $\pphi\in\mathbb{R}^{M}$ and the laser wavelength $\lambda$. Because the SM has no preferred direction with respect to $\pphi$, we assume for simplicity that the speckle field decorrelates isotropically in phase space. Under this assumption, estimating $N_{\rm HV}$ reduces to a packing problem of counting distinguishable points in the $M$-dimensional phase space at each wavelength.

We first consider the input phase channel. We assume that the phase shift at which the generated HVs decorrelate can be represented by $\Delta\phi \in \mathbb{R}$. Within a $2\pi$ period, each phase shifter then supports $n_\phi=\lfloor 2\pi/|\Delta\phi |\rfloor$ distinguishable levels. Because exchanging the two phase shifters of an MZM leaves its output amplitude unchanged and intensity detection is insensitive to a common phase offset of all shifters, the number of input-addressed HVs is
\begin{equation}
  N_{\rm input} \;=\; \frac{1}{n_\phi}\left[\frac{n_\phi(n_\phi+1)}{2}\right]^{M'},
  \label{eq:si_ninput}
\end{equation}
where $M'=M/2$ is the number of MZMs. Including the $N_\lambda$ wavelength addresses, the total number is
\begin{equation}
  N_{\rm HV} \;=\; N_{\lambda}N_{\rm input}.
  \label{eq:si_nhv}
\end{equation}

In our system, Fig.~2(b) in the main text gives $\Delta\phi\simeq 0.36\pi$ for $k=5$, so that $n_\phi=5$. With $M'=7$ and $N_\lambda\simeq 600$ (a $50\,\mathrm{pm}$ shift over $1530$--$1560\,\mathrm{nm}$ decorrelates the HVs), Eq.~(\ref{eq:si_nhv}) gives $N_{\rm HV}\simeq2\times10^{10}$. This estimate is sensitive to $n_\phi$: for $n_\phi=4$--$7$, $N_{\rm HV}$ ranges from $1.5\times10^{9}$ to $1.2\times10^{12}$. Importantly, $N_{\rm HV}$ grows exponentially with $M'$ irrespective of the value of $n_\phi$.

\section{Power budget and inference energy}\label{sec:power}

This note details the power and energy estimates quoted in the Discussion. We first give the estimated breakdown for the proof-of-concept system (\snref{sec:pof_sys}), then describe an architecture in which the $L$ class-HV similarities are evaluated in parallel (\snref{sec:parallel}), and finally estimate its inference energy and energy--delay product (EDP) on a common benchmark (\snref{sec:edp}). 

\subsection{Present experimental system}\label{sec:pof_sys}

The power budget of the system used for the experiments in the main text is the sum of four contributions, listed in Table~\ref{tab:power_present}. The laser power is the optical power delivered into the input port of the PIC. The modulator term is the electrical power dissipated by the $M=14$ thermo-optic phase shifters at the operating point required for a $\pi$ phase shift, $P_{\rm MZM} = V^2_{\pi}/R_h \approx 40.6\,\mathrm{mW}$ with $V_{\pi} \approx 3.12\,\mathrm{V}$ and $R_{h} \approx 240\,\Omega$. 
The DMD power is an upper bound on the power consumption of the DMD chip alone. It is obtained from the maximum supply currents: $P_{\rm DMD} =  3.6\,\mathrm{V}\times(650+350)\,\mathrm{mA}+8.75\,\mathrm{V}\times25\,\mathrm{mA}\approx3.8\,\mathrm{W}$~\cite{TI_DLP650LNIR}. The power consumption of the camera is estimated from a camera with the same sensor (Sony IMX991) and a two-stage thermoelectric cooler (QHYCCD QHY991). The camera is specified to consume $10\,\mathrm{W}$ at $50\%$ cooling power \cite{QHY991}.

The total power is estimated as
\begin{equation}
  P_{\rm total}
  = P_{\rm laser} + M P_{\rm MZM} + P_{\rm DMD} + P_{\rm camera}
  = 14.40\,\mathrm{W},
  \label{eq:si_power_present}
\end{equation}
of which the image sensor and DMD account for $13.8/14.40 \approx 96\%$.

\begin{table}[ht]
  \centering
  \caption{\textbf{Power breakdown of the proof-of-concept system.} $M=14$ addressed input channels.}
  \label{tab:power_present}
  \begin{tabular}{@{}llr@{}}
    \toprule
    Component & Specification & Power \\
    \midrule
    Laser  & optical power, $\lambda=1550\,\mathrm{nm}$
                         & $30\,\mathrm{mW}$ \\
    Modulator array      & $M\times P_{\rm MZM}$, $P_{\rm MZM} \approx 40.6\,\mathrm{mW}$
                         & $0.57\,\mathrm{W}$ \\
    DMD      & $1280\times800$, 10.752 kHz
                         & $3.8\,\mathrm{W}$ \\
    InGaAs camera & $640\times512$, TEC-cooled, 133~fps
                         & $10\,\mathrm{W}$ \\
    \midrule
    \textbf{Total}       & & $\bm{14.40\,\mathrm{W}}$ \\
    \bottomrule
  \end{tabular}
\end{table}

\subsection{Architecture for parallel similarity evaluation}\label{sec:parallel}

\begin{figure}[H]
  \centering
  \includegraphics[width=\textwidth]{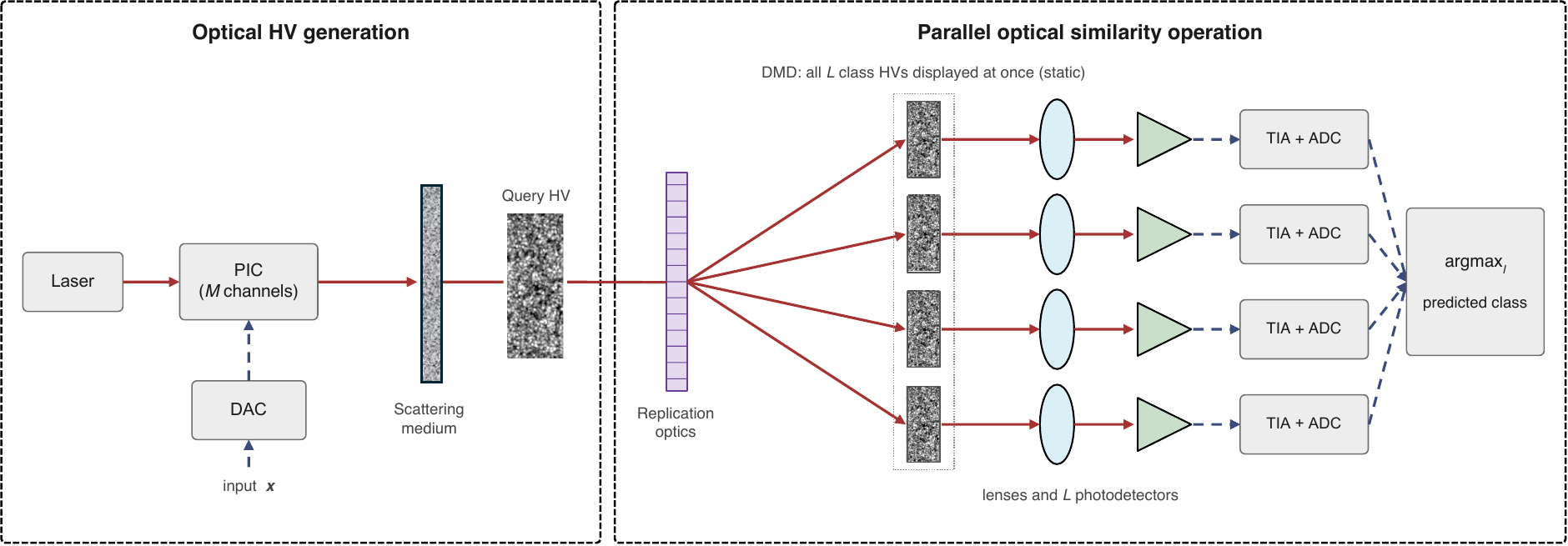}
\caption{\textbf{Architecture for parallel optical similarity evaluation.} 
Replication optics generate $L$ copies that illuminate disjoint DMD regions displaying the class HVs $\cc_\ell$. Separate detection of each region enables simultaneous evaluation of the similarity to all $L$ classes in a single optical shot.}
  \label{fig:si_parallel}
\end{figure}

The experimental system can be extended to parallel similarity evaluation with two modifications (Supplementary Fig.~\ref{fig:si_parallel}). First, the image sensor is replaced by $L$ photodetectors with transimpedance amplifiers (TIAs) and analog-to-digital converters (ADCs). Second, the query speckle pattern is replicated into $L$ copies, which illuminate disjoint regions of a DMD displaying the class HVs $\cc_\ell$. This enables all $L$ inner products to be evaluated in a single optical shot. Fan-out optics such as microlens arrays \cite{Wang_NP:2023} and Dammann gratings \cite{DAMMANN1971312,Ma_Dammanngrating:23} are candidate implementations, although replicating diffraction-limited speckle patterns presents an additional challenge. Alternatively, disjoint regions of the same speckle field can be used. Both configurations have the same computational scaling.

Because the class HVs are fixed after training, the DMD can display a static pattern, and its refresh rate does not limit the inference throughput. A fixed lithographic mask could therefore replace the DMD. The number of classes evaluated in parallel is bounded by the mirror count. For example, a DMD with the mirror count of the DLP9500~\cite{TI_DLP9500} provides $\lfloor 2.0736\times10^{6}/D\rfloor$ regions, corresponding to a maximum of $506$ classes at $D=4096$ and $207$ at $D=10^{4}$.

\subsection{Inference energy and energy--delay product}\label{sec:edp}

Following the evaluation methodology of PhotoHDC \cite{Fayza:2026}, we adopt the same device references for the modulators, DACs, and ADCs. We consider the ISOLET dataset, with $M=617$ input features, $L=26$ classes, and $D=4096$. The input rate is $f_{\rm in}=5\,\mathrm{GHz}$ at 4-bit resolution, and the readout uses 10-bit ADCs.

The power assumptions are as follows. Each driven phase shifter consumes $P_{\rm MZM}=5.7\,\mathrm{mW}$, corresponding to one arm of the $11.3\,\mathrm{mW}$ push--pull RF drive power reported for a silicon Mach--Zehnder modulator \cite{Akiyama:2012}. 
The input DAC energy is estimated by scaling the energy of the 14-bit converter \cite{Huang_DAC:2020}, assuming that the DAC energy scales as $2^b$ with the nominal resolution $b$: $E_{\rm DAC}=17.7\,\mathrm{pJ}\times2^{4}/2^{14} =17.3\,\mathrm{fJ}$ per sample. The corresponding power consumption is $P_{\rm DAC}=E_{\rm DAC}f_{\rm in}=86\,\mu\mathrm{W}$ per channel. The output ADCs use the $10$-bit, $5\,\mathrm{GS/s}$ device of Ref.~\cite{Guo_ADC:2019} used at its native resolution, $P_{\rm ADC}=29\,\mathrm{mW}$. For the TIAs, we take $P_{\rm TIA}=8\,\mathrm{mW}$~\cite{Wang_TIA:2013}. 
The DMD is assumed to have the same number of micromirrors as the Texas Instruments DLP9500 \cite{TI_DLP9500}. Its power consumption is $P_{\rm DMD}=4.4\,\mathrm{W}$ \cite{TI_DLP9500}. The laser power is assumed to be $P_{\rm laser}=30\,\mathrm{mW}$, as in the present experiment.

The total power for reconfigurable prototypes using the DMD is
\begin{equation}
  P_{\rm total}
  = P_{\rm laser} + P_{\rm DMD}
  + M\!\left(P_{\rm DAC}+P_{\rm MZM}\right)
  + L\!\left(P_{\rm ADC}+P_{\rm TIA}\right)
  = 8.96\,\mathrm{W},
  \label{eq:si_power_future}
\end{equation}
with the breakdown given in Table~\ref{tab:power_future}.
During inference, the displayed DMD patterns are fixed. If the DMD is replaced by a fixed lithographic mask, $P_{\rm DMD}$ is omitted.

In the proposed scheme, the entire $D$-dimensional HV and all $L$ similarities
are obtained from a single optical propagation. Assuming that successive
samples are processed at the input rate $f_{\rm in}$, the processing time per sample is $T_D = 1/f_{\rm in}=0.2\,\mathrm{ns}$. The energy per inference is estimated as 
\begin{equation}
  E_{\rm inf} = P_{\rm total}T_D = 1.79\,\mathrm{nJ/sample}.
\end{equation}
The energy--delay product (EDP), defined here as $\mathrm{EDP} = E_{\rm inf}T_D$ with the processing time (inverse throughput) $T_D$, is estimated as
\begin{equation}
  \mathrm{EDP} = 3.58\times10^{-19}\,\mathrm{J\,s}.
  \label{eq:si_edp} 
\end{equation}
Table~\ref{tab:si_compare} compares these values with the electro-photonic accelerator PhotoHDC \cite{Fayza:2026} and with a compute-in-memory (CiM) accelerator evaluated on the same benchmark. In PhotoHDC, the encoding of one sample is instead distributed over successive cycles of a photonic array of fixed size, which occupies the array for an estimated $T_D = 8.7\,\mathrm{ns}$ per sample. 
The CiM figure is the value reported in Ref.~\cite{Fayza:2026}, obtained by modeling the same HDC workload on $128\times 128$ ReRAM subarrays at 22 nm with the DNN+NeuroSim framework \cite{Peng_NeuroSim:2019} at a $1~\mathrm{GHz}$ clock.

\paragraph*{Limits of the comparison}

\emph{Free-space versus integrated implementation.}
The proposed system is a free-space optical arrangement, whereas both comparison platforms are integrated. Its projected advantage partly arises from carrying $D$ spatial modes without a per-mode fabricated element. A fair comparison would require an on-chip implementation of the proposed architecture.

\emph{Scope of the estimate.}
Equation~\eqref{eq:si_power_future} covers HV generation and similarity evaluation only. It excludes electronic bit-plane selection, binding, bundling, and class-HV construction, as well as optical insertion losses beyond the fixed laser power. The estimate assumes static class HVs and therefore applies to inference with a trained model, not online learning.

\begin{table}[H]
  \centering
  \caption{\textbf{Projected power breakdown for parallel similarity evaluation.} $M=617$, $L=26$, $D=4096$, and $f_{\rm in}=5\,\mathrm{GHz}$. The DMD is included for the reconfigurable configuration and omitted for the fixed-mask configuration.}  \label{tab:power_future}
  \begin{tabular}{@{}llrr@{}}
    \toprule
    Component & Scaling & Power & Share \\
    \midrule
    Laser                  & ---                          & $0.03\,\mathrm{W}$ & $0.3\%$ \\
    DMD ($1920\times1080$) & ---                          & $4.40\,\mathrm{W}$ & $49.1\%$ \\
    DAC (4-bit)            & $M\times86\,\mu\mathrm{W}$   & $0.05\,\mathrm{W}$ & $0.6\%$ \\
    Modulators             & $M\times5.7\,\mathrm{mW}$    & $3.52\,\mathrm{W}$ & $39.2\%$ \\
    ADC (10-bit)           & $L\times29\,\mathrm{mW}$     & $0.75\,\mathrm{W}$ & $8.4\%$ \\
    TIA ($8.4\,\mathrm{GHz}$)& $L\times8\,\mathrm{mW}$    & $0.21\,\mathrm{W}$ & $2.3\%$ \\
    \midrule
    \textbf{Total, reconfigurable prototypes (DMD)}         &                              & $\bm{8.96\,\mathrm{W}}$ & \\
    \textbf{Total, fixed prototypes (mask)}         & \text{DMD term removed}                             & $\bm{4.56\,\mathrm{W}}$ & \\
    \bottomrule
  \end{tabular}
\end{table}

\begin{table}[H]
  \centering
  \caption{\textbf{Comparison of inference energy and energy--delay product on the ISOLET dataset.} }
  \label{tab:si_compare}
  \begin{tabular}{@{}lrrrr@{}}
    \toprule
    Platform & Rate $f_{\rm in}$ & $E_{\rm inf}$ (nJ/sample) & EDP (J\,s) & EDP ratio \\
    \midrule
    CiM                     & $1\,$GHz & ---   & $1.25\times10^{-12}$ & $3.5\times10^{6}$ \\
    PhotoHDC\cite{Fayza:2026} & $5\,$GHz & $90.1$ & $7.82\times10^{-16}$ & $2.2\times10^{3}$ \\
    \textbf{This work (projected)} & $5\,$GHz & $\bm{1.79}$ & $\bm{3.58\times10^{-19}}$ & --- \\
    \bottomrule
  \end{tabular}
\end{table}

\section{Statistical quality of the generated HVs}\label{sec:stats}
We assess whether the generated HVs reproduce the statistics of ideal independent random vectors. The sample comprises $N_{\lambda}=601$ HVs, one per wavelength over $1530$--$1560\,\mathrm{nm}$ in $50\,\mathrm{pm}$ steps, at $D=327{,}680$ and $k=5$, mapped to bipolar form. Table~\ref{tab:si_stats} lists the measured values together with the ideal values expected for independent random bipolar HVs.

Six of the seven quantities agree with the ideal value within about $2\%$. The only exception is the per-vector component mean, whose standard deviation is $3.16$ times the ideal value. This deviation is attributed to the inhomogeneous speckle illumination, which produces intensity variations across the detection positions.

\begin{table}[H]
  \centering
  \caption{\textbf{Statistical tests on the generated HVs.}
  $D=327{,}680$, $k=5$, $N_{\lambda}=601$. All measured and ideal values are in
  units of $10^{-3}$. Ideal values are those of independent random bipolar HVs:
  $1/\sqrt{D}$, except $\sqrt{2/(\pi D)}$ for the adjacent-wavelength mean $|C|$
  and $1/\sqrt{N_{\lambda}}$ for the per-pixel mean. s.d.\ denotes the standard
  deviation.}
  \label{tab:si_stats}
  \begin{tabular}{@{}llrrr@{}}
    \toprule
    Test & Quantity & Measured & Ideal & Ratio \\
    \midrule
    Quasi-orthogonality & s.d.\ of pairwise cosine similarity & $1.745$ & $1.747$ & $0.999$ \\
                        & mean $|C|$, adjacent wavelengths & $1.421$ & $1.394$ & $1.020$ \\
    Balance             & s.d.\ of per-HV component mean & $5.51$ & $1.747$ & $\bm{3.16}$ \\
                        & s.d.\ of per-pixel mean over HVs & $41.0$ & $40.79$ & $1.005$ \\
    Closure under binding & s.d.\ among two-fold products & $1.755$ & $1.747$ & $1.005$ \\
                        & s.d.\ among three-fold products & $1.742$ & $1.747$ & $0.997$ \\
                        & s.d.\ between products and factors & $1.769$ & $1.747$ & $1.013$ \\
    \bottomrule
  \end{tabular}
\end{table}

\clearpage
\bibliographystyle{unsrtnat}
\bibliography{samples_ver4}